\documentclass[twocolumn]{aa} % for a paper on 2 column  
\usepackage{txfonts}
\usepackage{natbib}
\usepackage{lscape}                                % Tables in landscape

\usepackage{color}
\usepackage{graphicx}
\usepackage{longtable}
\newcommand{\pyneb} {{\sc PyNeb}}

\newcommand{\elecd}{$n_{\rm e}$}
\newcommand{\te}{$T_{\rm e}$}
\newcommand{\hb}{H$\beta$}
\newcommand{\ha}{H$\alpha$}

\newcommand{\foiii}{[O\,{\sc iii}]}
\newcommand{\foii}{[O\,{\sc ii}]}
\newcommand{\fsii}{[S\,{\sc ii}]}
\newcommand{\fsiii}{[S\,{\sc iii}]}
\newcommand{\fnii}{[N\,{\sc ii}]}
\newcommand{\fariii}{[Ar\,{\sc iii}]}
\newcommand{\fariv}{[Ar\,{\sc iv}]}
\newcommand{\farv}{[Ar\,{\sc v}]}
\newcommand{\fcliii}{[Cl\,{\sc iii}]}
\newcommand{\fneiii}{[Ne\,{\sc iii}]}
\newcommand{\fnev}{[Ne\,{\sc v}]}
\newcommand{\hi}{H\,{\sc i}}
\newcommand{\hii}{H\,{\sc ii}}
\newcommand{\hei}{He\,{\sc i}}
\newcommand{\heii}{He\,{\sc ii}}

\newcommand{\mc}{\multicolumn}
\newcommand{\nodata}{---}

\begin{document}

\title{The PNe and {\hii} regions in NGC\,6822 revisited. Clues to AGB nucleosynthesis.} 

\author{
Jorge Garc\'ia-Rojas\inst{1,2}, 
 Miriam Pe\~na\inst{3}, Sheila Flores-Dur\'an\inst{3}, and Liliana Hern\'andez-Mart{\'\i}nez\inst{3,4} } 
\offprints{} 
\institute{
1 Instituto de Astrof\'isica de Canarias, E-38205 La Laguna, Tenerife, Spain\\ 
2 Universidad de La Laguna, Dept. Astrof\'{\i}sica. E-38206 La Laguna, Tenerife, Spain \\
3 Instituto de Astronom\'ia, Universidad Nacional Aut\'onoma de M\'exico,Apdo. Postal 70-264, M\'ex. D. F., 04510 M\'exico\\
4 Instituto de Ciencias Nucleares, Universidad Nacional Aut\'onoma de M\'exico,  Apdo. Postal 70-543, M\'ex. D. F., 04510 M\'exico\\
\email{jogarcia@iac.es}  }
\date{Received 28 August 2015; accepted 18 October 2015} 

\titlerunning{Abundances in PNe in NGC\,6822.}

\authorrunning{Garc{\'\i}a-Rojas et al.} 

%----------Abstract-----------------------

\abstract 
{}
{The chemical behaviour of an ample sample of PNe in NGC6822 is analysed. } 
{Spectrophotometric data of 11 PNe and two {\hii} regions were obtained with the OSIRIS spectrograph attached to the Gran Telescopio Canarias. Data  for other 13 PNe and three {\hii} regions were retrieved from the literature. Physical conditions and chemical abundances of O, N, Ne, Ar, and S were derived in a consistent way for 19 PNe and 4 {\hii} regions.    } {Abundances in the PNe sample are widely distributed showing 12+log (O/H) from  7.4 to 8.2 and 12+log (Ar/H) from 4.97 to 5.80. Two groups of PNe can be differentiated: one  old with low metallicity (12+log (O/H) $<$ 8.0 and 12+log (Ar/H) $<$ 5.7) and another younger one with metallicities similar to the values for {\hii} regions. The old objects are  distributed in a larger volume than the young ones.  An important fraction of PNe (over 30\%) was found to be highly N-rich (Peimbert Type I PNe). Such PNe occur at any metallicity. In addition, about 60\%  of the sample presents high ionization (He$^{++}$/He $\geq$ 0.1), possessing a central star with effective temperature higher than 100~000 K. Possible biases in the sample are discussed. From comparison with stellar evolution models by Karakas (2010) and Fishlock et al. (2014) of the observed  N/O abundance ratios, our PNe should have had initial masses that are lower than 4 M$_\odot$, although if the comparison is made with Ne vs. O abundances, the initial masses  should have been lower than 2 M$_\odot$. It appears that these models of stars of  2$-$3 M$_\odot$ are producing too much $^{22}$Ne in the stellar surface at the end of the AGB. On the other hand, the comparison with another set of stellar evolution models 
with a different treatment of convection and on the assumptions about the overshoot of the convective core during the core H-burning phase, provided there is reasonable agreement between the observed and predicted N/O and Ne/H ratios if initial masses of more massive stars are about 4 M$_{\odot}$.} 
{}


\maketitle 

\section{Introduction\label{sec:intro}}
The barred irregular galaxy NGC\,6822 (IBs(m), also known as DDO209), is one of the most interesting  galaxies in the Local Group. Half way between the Milky Way and Andromeda, \citep[reported a distance modulus of 23.31$\pm$0.02]{gierenetal06}, it presents several very interesting characteristics that have made it the subject of intense studies.

NGC\,6822 contains different populations with different spatial distributions and kinematics. Its optical structure is dominated by a bar that is 8$'$ long (about 1.1 kpc), oriented almost N-S, and it contains a huge disk of {\hi} of about 6$\times$13 kpc at a position angle (P.A.) of 130$^\circ$ centred on the optical centre \citep{deblokwalter00, weldrakeetal03}. In addition, a huge spheroidal stellar distribution with its long axis at P.A. of 64.5$^\circ$ (almost perpendicular to the {\hi} disk), composed mainly of intermediate-age stars, has been found \citep{letarteetal02, battinellietal06}.  Interestingly, the {\hi} disk and the stellar spheroid  do not share the same kinematics. The stellar spheroid has a systemic radial velocity v$_{sys} = -32.9$ km s$^{-1}$,  and it rotates around its minor axis with  radial velocities extending from 20 to $-70$ km s$^{-1}$ \citep{demersetal06}, while the {\hi} disk  shows a systemic velocity v$_{sys}$ = $-$57 km s$^{-1}$ and rotates around an axis almost orthogonal to the main axis of the spheroid, with velocities extending from -51 km s$^{-1}$ to 100 km s$^{-1}$ \citep{deblokwalter06}. A recent study of several stellar clusters in NGC\,6822 \citep{hwang14} reveals  that the clusters seem to show completely different kinematics with a mean radial velocity of $-88  \pm 22.7$ km s$^{-1}$.

In recent years we have been analysing the PN population in this galaxy and other galaxies and its relation with other objects (e.g., Pe\~na et al. 2007 for NGC\,3109; Hern\'andez-Mart{\'\i}nez et al. 2009, 2011, for NGC\,6822;  Stasi\'nska et al. 2013 for NGC\,300).

Chemically, NGC\,6822 is a metal-poor galaxy similar to the SMC, with an interstellar medium (ISM) abundance of about 0.2 $Z_\odot$ \citep{richermccall07}. \citet{hdezmartinezpena09} discovered 26 PN candidates in NGC\,6822, and \citet{hdezmartinezetal09} analysed the chemical abundances of a limited sample of them (11 objects), showing that the studied PNe belong to two groups of different ages (1-3 Gyr and 3-9 Gyr) and different abundances, the first one with a composition similar to the present ISM and the older one with abundances  a factor of 2 lower. By studying abundances in {\hii}  regions, they also found that the ISM located in the 2 kpc around the centre is chemically homogeneous, showing 12+ log (O/H) = 8.06$\pm$0.04. 

\citet{carigietal06} constructed chemical evolution models that fit the photometric properties of NGC\,6822 computed by \citet{wyder01,wyder03} and derived a robust star formation history (SFH) for this galaxy. Detailed galactic chemical evolution models were computed by Hern\'andez-Mart\'{\i}nez et al. (2009, 2011) using PNe and {\hii}  region abundances as constraints with no conclusive results. One of the two best models reproduces the abundances of O, Ne, S, Ar, and C derived from collisionally excitation lines (CELs) if the most massive star in the initial mass function has a mass M$_{up}$= 40 M$_\odot$. The other model reproduces the abundances obtained from recombination lines (RLs, which are about a factor of 2 higher), if a M$_{up}$= 80 M$_\odot$ is assumed. This is a direct effect of the abundance discrepancy problem, a still unresolved problem for photoionized nebulae, which consists in the well known fact that abundances obtained from RLs are systematically larger than those obtained from CELs of the same ion \citep[e.g.,][]{garciarojasesteban07, liu12, estebanetal09}. This effect is generally parametrized by the abundance discrepancy factor (ADF), which is defined as the ratio between abundances obtained from RLs and CELs. The ADF is usually between 1.5 and 3, but in PNe it has a significant tail that extends to much higher values, up to 120 \citep{corradietal15a}. By analysing high-resolution spectra, \citet{floresduranetal14} demonstrate that the kinematics of PNe is closer to the one shown by the stellar spheroid than to the {\hi} disk, thus PNe would belong to the intermediate age population. 

In this work we revisit the PN population in order  to analyse the chemical behaviour of a larger sample of objects. New data were obtained from observations with the Gran Telescopio Canarias (GTC).
In \S~\ref{sec:obs} we present the observations and data reduction. Section~\ref{sec:PN} is devoted to discussing the true nature of the sample.  In \S~\ref{sec:ionic} the physical conditions (electron temperatures and densities) and ionic abundances of a wide sample of objects are calculated. In \S~\ref{sec:totalab} the total abundances are derived and discussed in comparison with stellar evolution models from the literature. Our results are presented in \S~\ref{sec:discuss}, and the final summary is found in \S~\ref{sec:summary}.

\section{Observations and data reduction
\label{sec:obs}}

Long-slit spectroscopy was performed with the OSIRIS spectrograph attached to the GTC, which was in service mode during 2014\footnote{GTC programme number is GTC5-14AIACMEX, and 6 hrs of observing time were awarded. The observing time was divided in 6 observing blocks (OB), one hour each, of which 45 min were used for scientific observations}. Three exposures of 900 s each were obtained for each slit position. A binning of 2$\times$2 was used.
Grism R1000B  (IR\_G3.2)  was employed to cover from 3630 to 7500 \AA\ with an effective resolution of $\sim$1000 at 7510 \AA. This wavelength range and resolution are well suited to detecting the main plasma diagnostic line ratios, which are useful for electron  temperature and electron density determinations in the nebula, which are crucial for an accurate ionic abundance determination.
 
The slit size was 6.8 arcmin long by 1.5 arcsec wide. Slits were oriented in a mode where two or more objects were observed simultaneously in each observing block (OB) to save exposure time.  The log of observations is presented in Table 1, where the slit position angle and observed objects in  each OB  
 are included. Because PNe are faint objects, most of the time a blind offset from a nearby field star was needed to place the objects in the slit.  In total we observed 11 PN candidates and two faint {\hii} regions. The effect of atmospheric differential refraction was minimized by observing NGC\,6822 close to culmination, which translates into slit orientations that are very close to the parallactic angle (differences less than 20-30$^{\circ}$). This was not the case for OB2 where slit position was more than 60$^{\circ}$ away from parallactic angle; however, spectra of PN\,25 and PN\,26 were not useful for our purposes, probably owing to large slit losses.

\begin{table}
\begin{tiny}
\caption{Log of observations with GTC$^a$}
\label{lines}
\begin{tabular}{lllll}
\noalign{\hrule} \noalign{\vskip3pt}
Block& Date & Slit  P.A.$^b$ & A.M. & Objects$^c$, night condition\\
\noalign{\smallskip} \noalign{\hrule} \noalign{\smallskip}
OB1 & 2014-06-02 &  $-$26.4$^\circ$ & 1.387 & PN\,2, PN\,8, dark clear\\
OB2 & 2014-06-03 &  +54.4$^\circ$ & 1.397 & PN\,25, PN\,26, dark spectr.\\
OB3 & 2014-06-23 &  $-$10.3$^\circ$&1.502 & PN\,23, PN\,24, dark clear\\
OB4 & 2014-06-24 &   +1.3$^\circ$ & 1.380 &PN\,11, PN\,15, dark clear\\
OB5 & 2014-06-27 &  $-$0.35$^\circ$ & 1.405 & PN\,9, PN\,22, dark phot.\\OB6 & 2014-06-27 &   +3.8$^\circ$& 1.389 & PN\,13, H\,II, H\,III, dark phot.\\
\noalign{\smallskip} \noalign{\hrule} \noalign{\smallskip}
\end{tabular}
\begin{description}
\item[$^{\rm a}$] Exposure time was always 3$\times$900 s, for each OB.
\item[$^{\rm b}$] Slit size was always 6.8 arcmin $\times$ 1.5 arcsec.
\item[$^{\rm c}$] PN names as in \citet{hdezmartinezpena09}.
\end{description}
\end{tiny}
\end{table}

\begin{figure*}
\caption{Calibrated spectra of PNe and {\hii} regions observed with GTC. Flux in units of 10$^{-16}$ erg cm$^{-2}$ s$^{-1}$ A$^{-1}$.\label{spectra}}
\includegraphics[width=0.8\columnwidth]{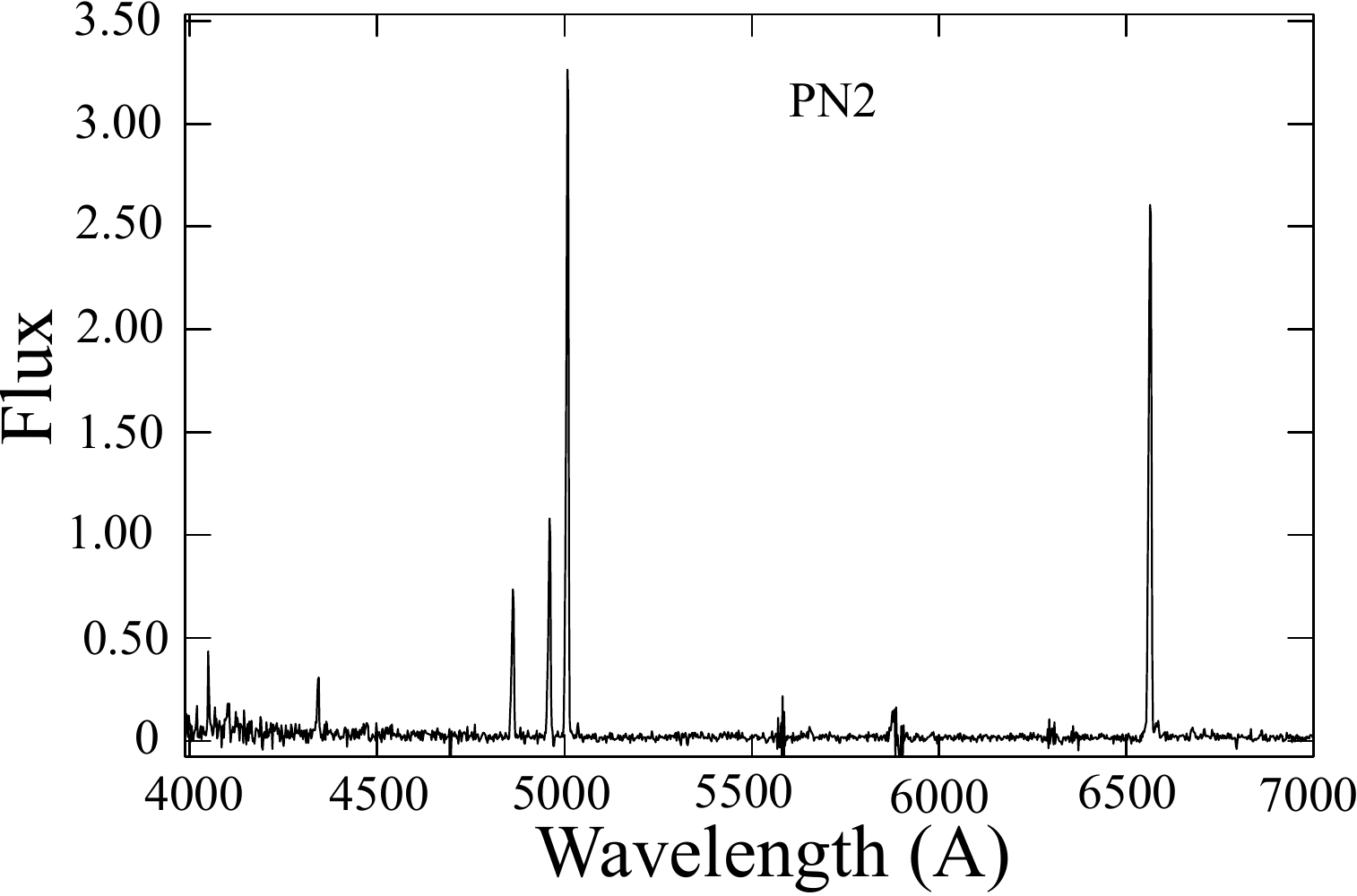}
\includegraphics[width=0.8\columnwidth]{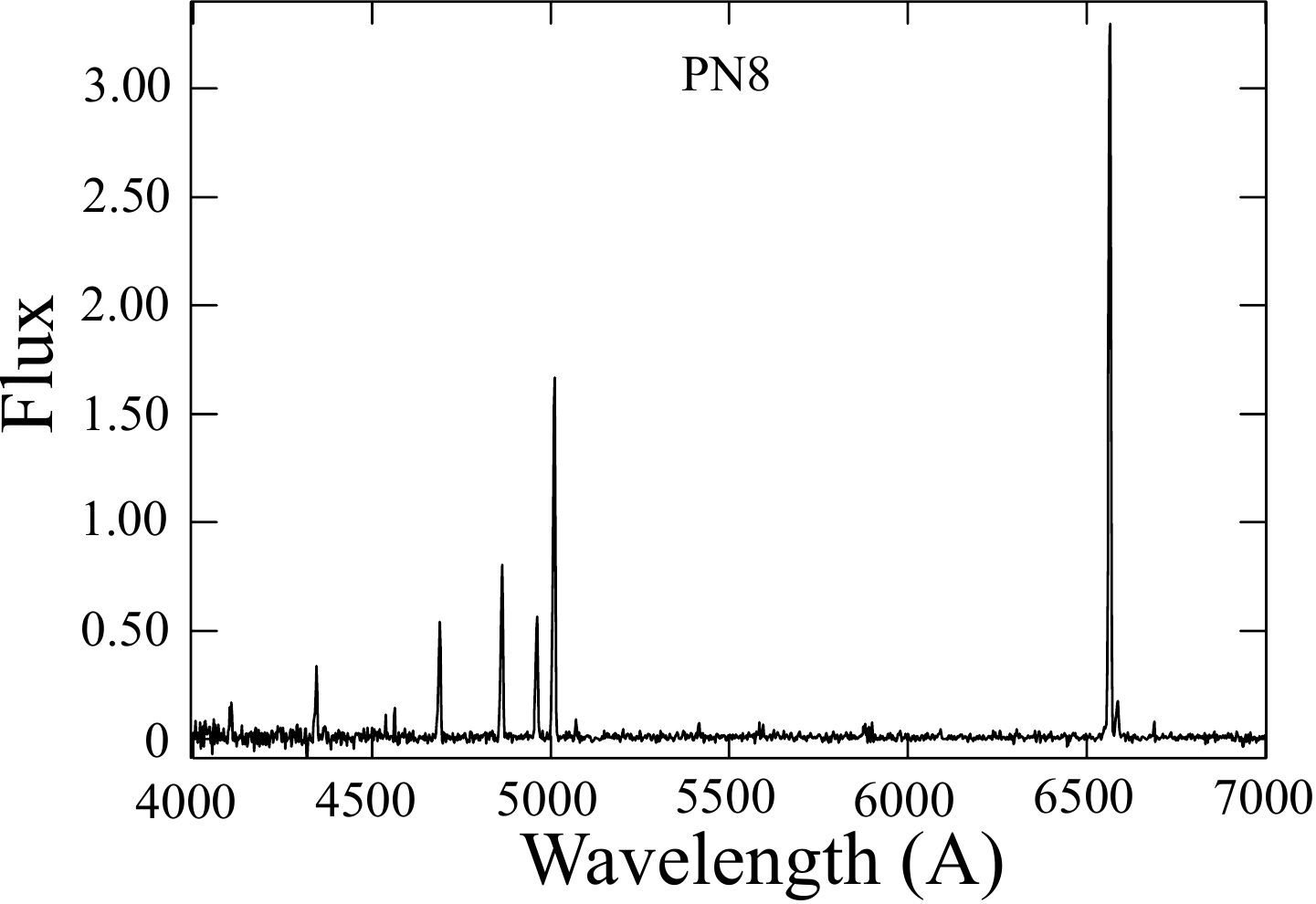}

\includegraphics[width=0.8\columnwidth]{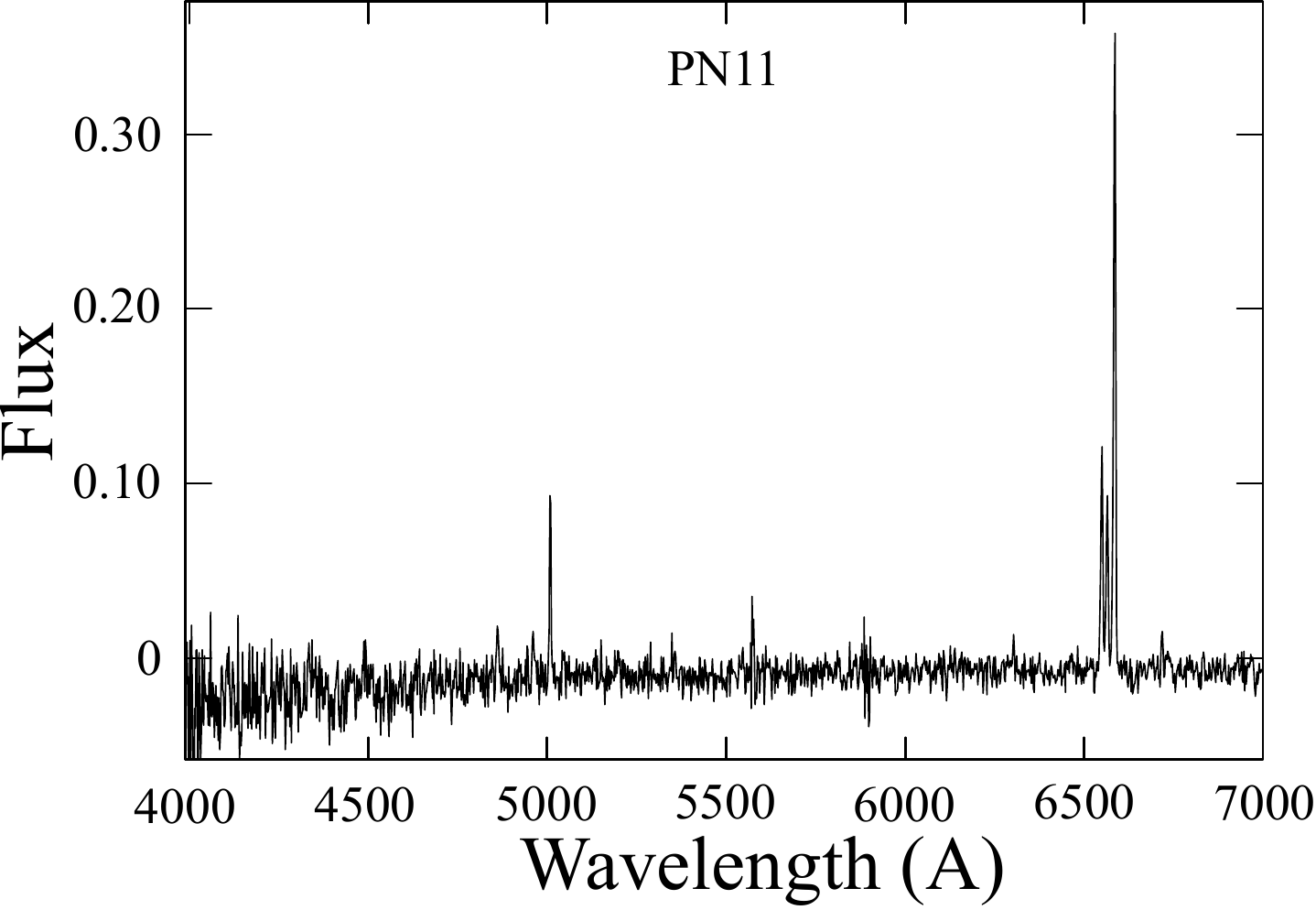}
\includegraphics[width=0.8\columnwidth]{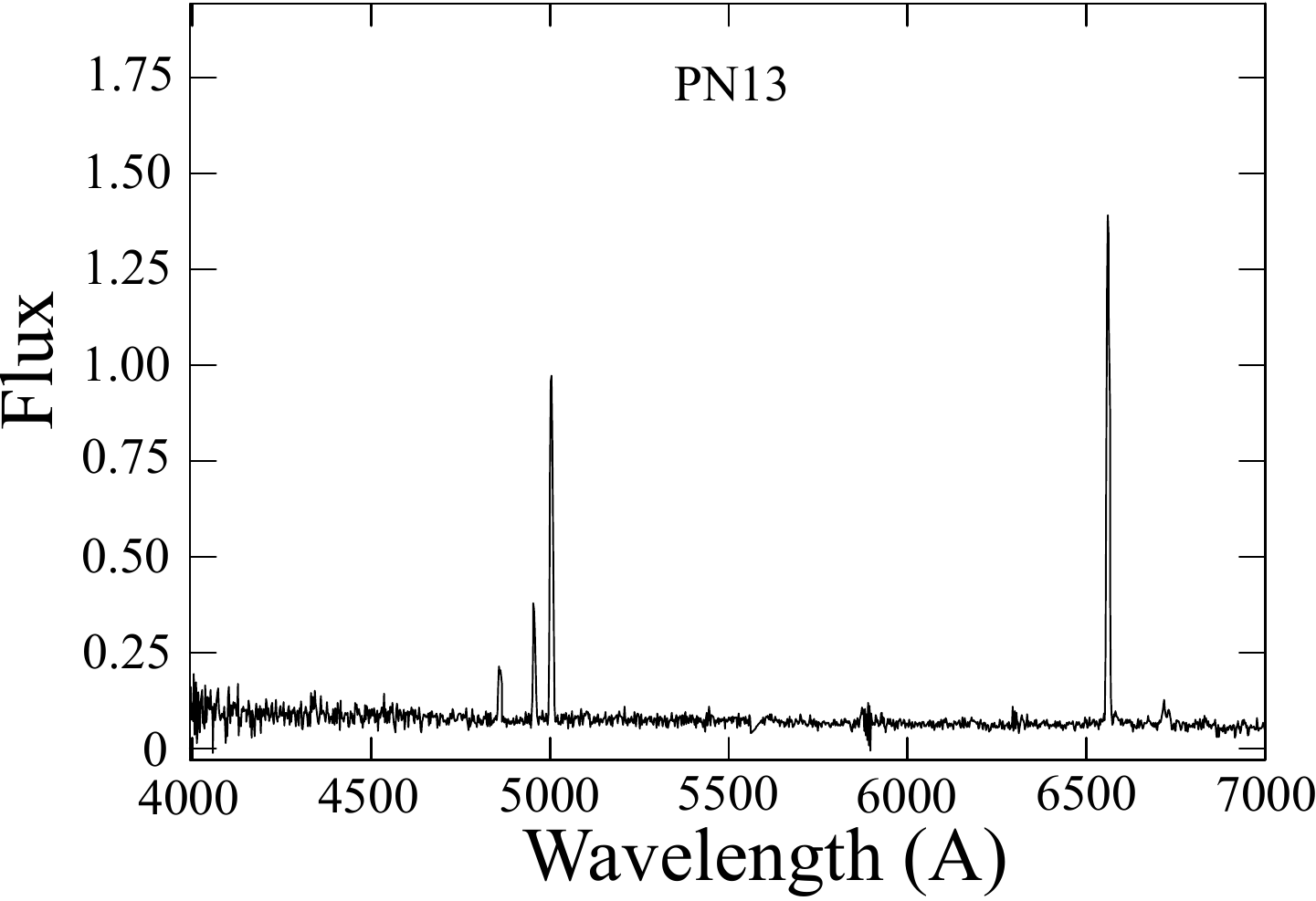}

\includegraphics[width=0.8\columnwidth]{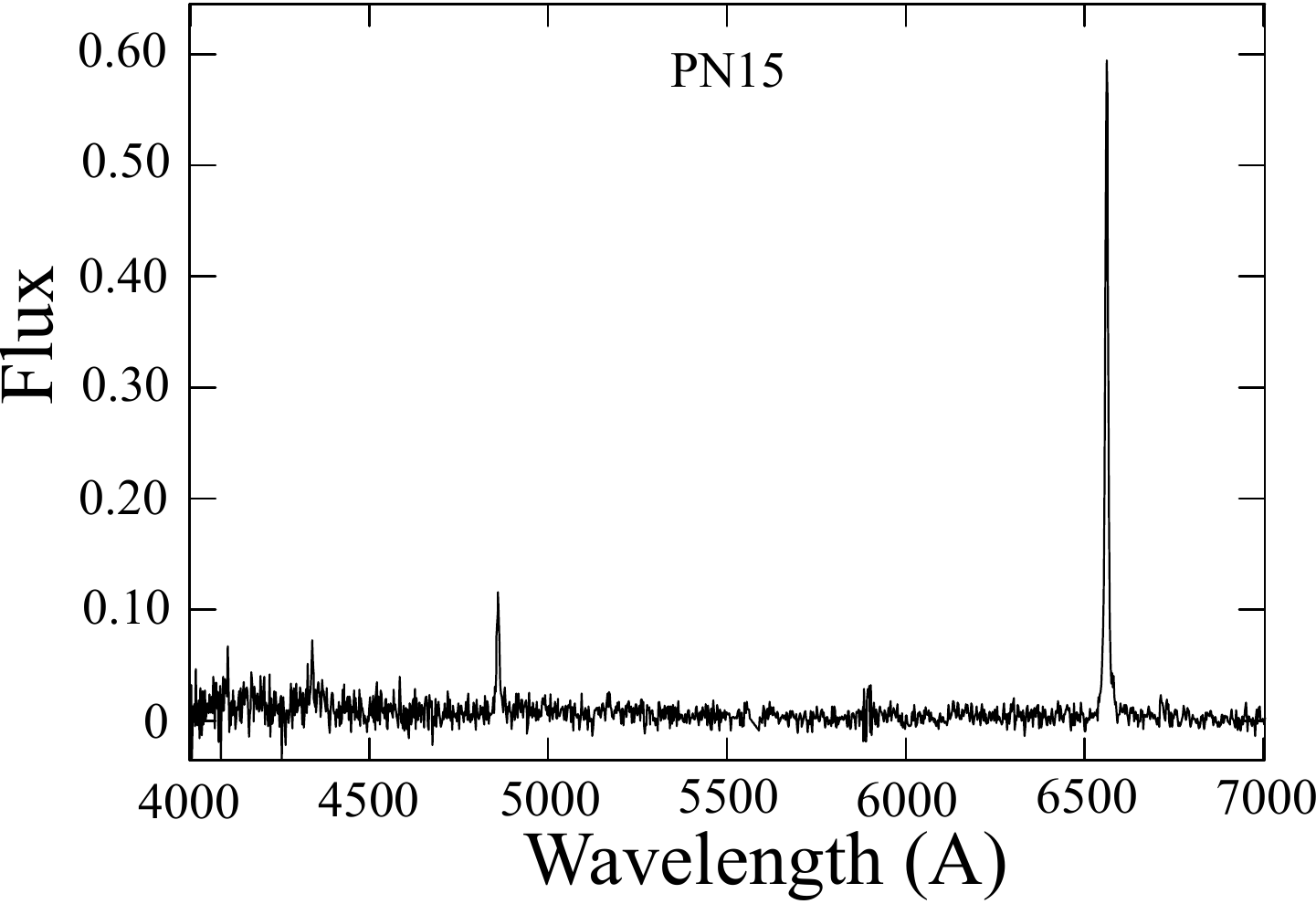}
\includegraphics[width=0.8\columnwidth]{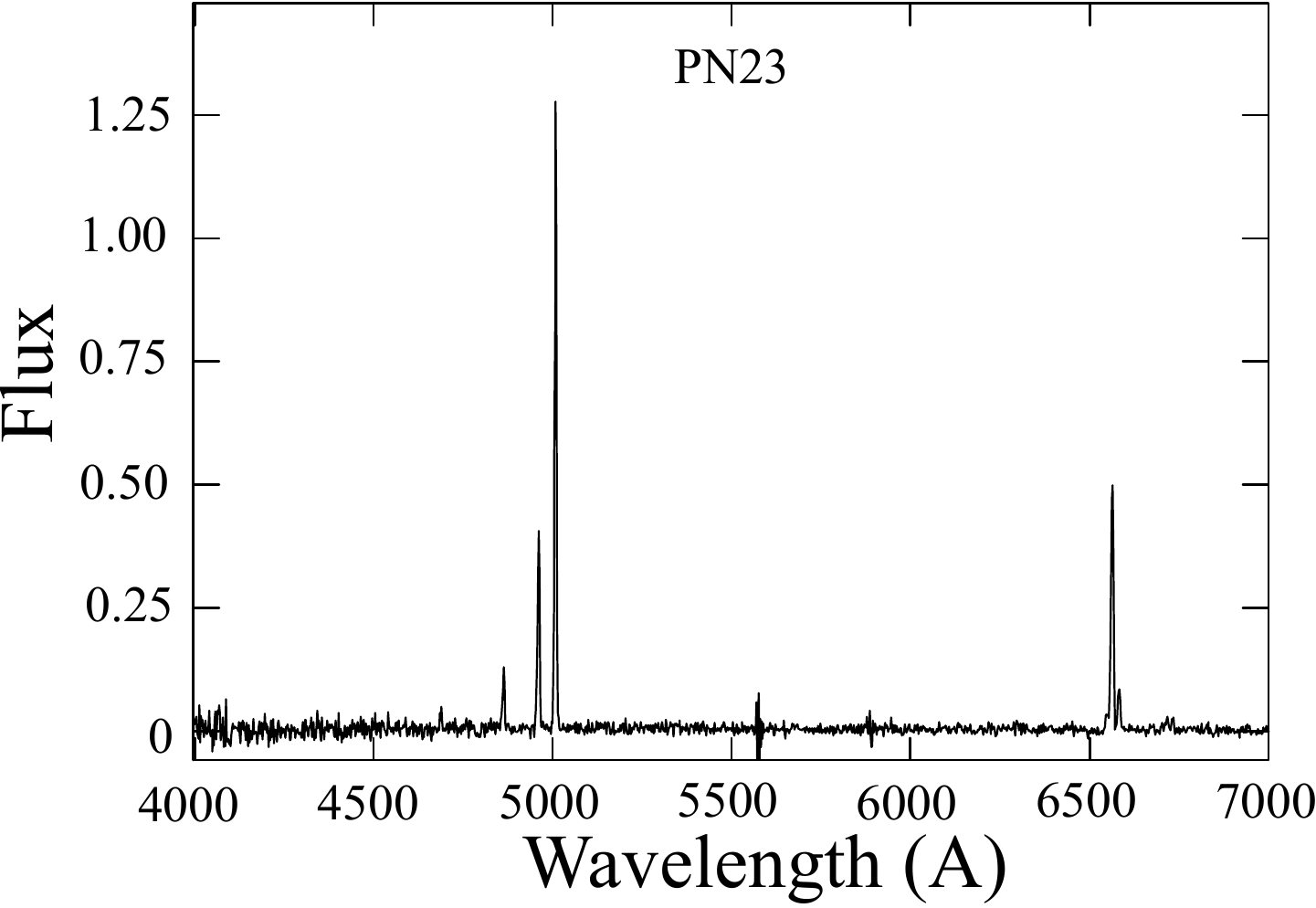}

\includegraphics[width=0.8\columnwidth]{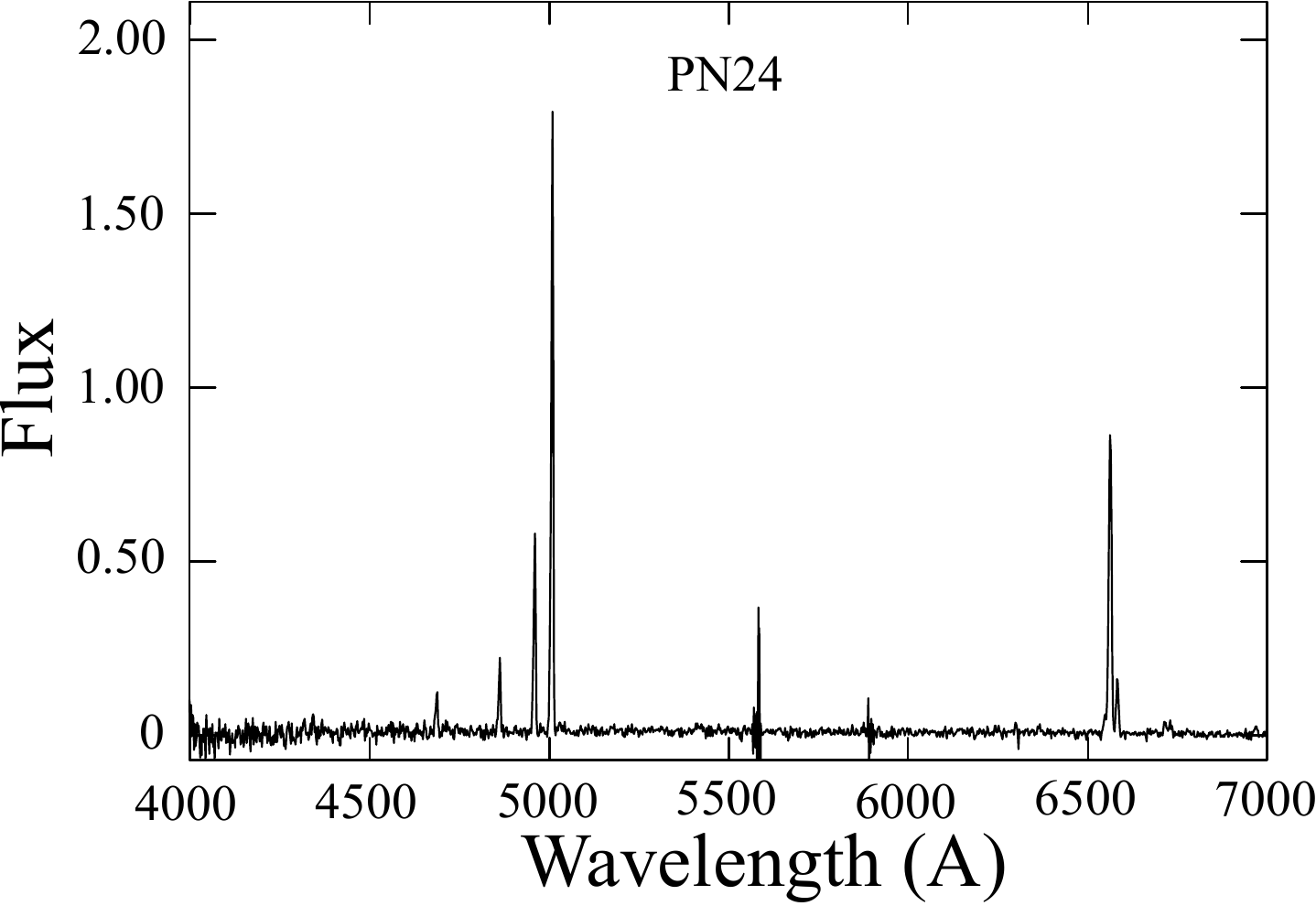}
\includegraphics[width=0.8\columnwidth]{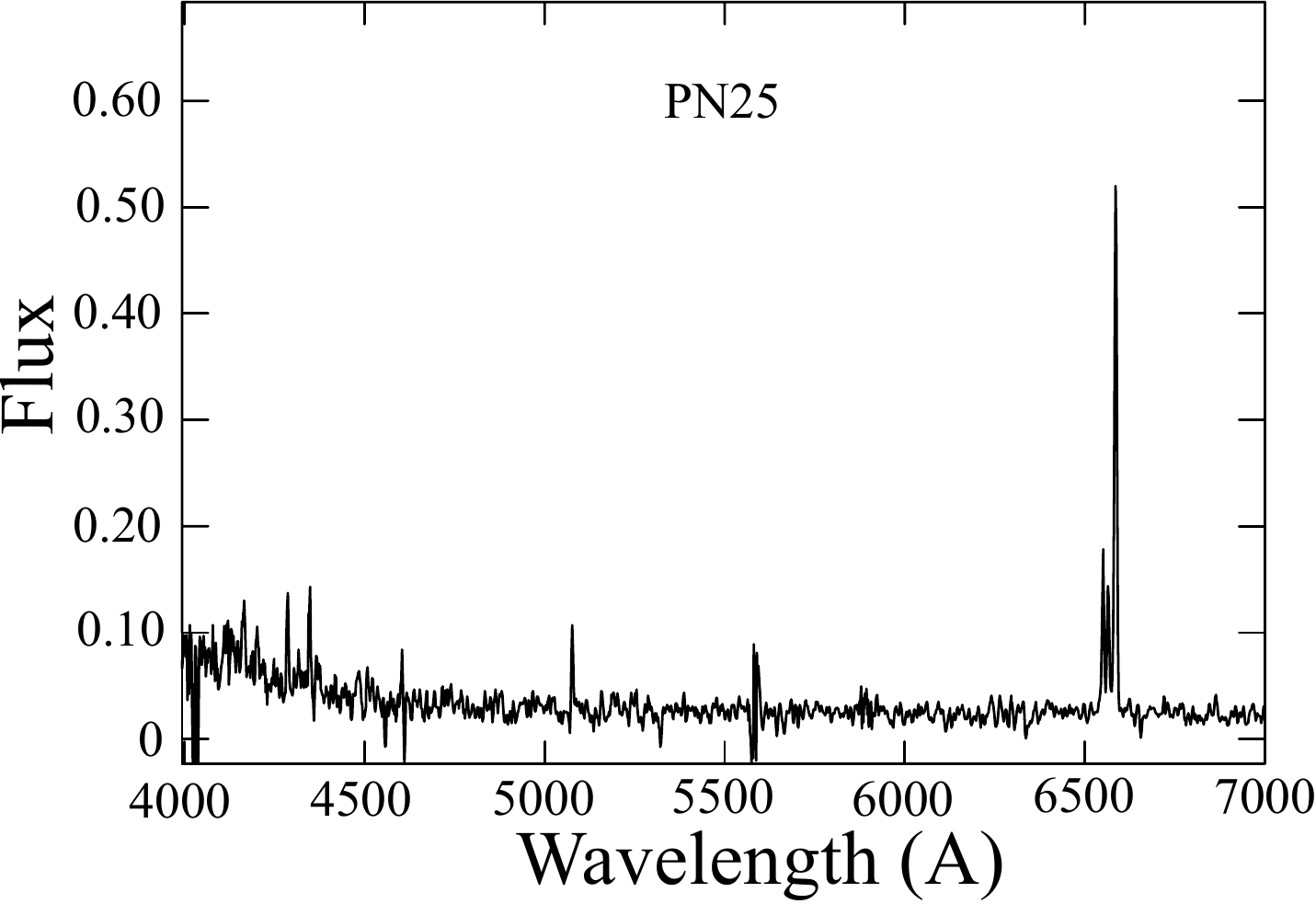}

\includegraphics[width=0.82\columnwidth]{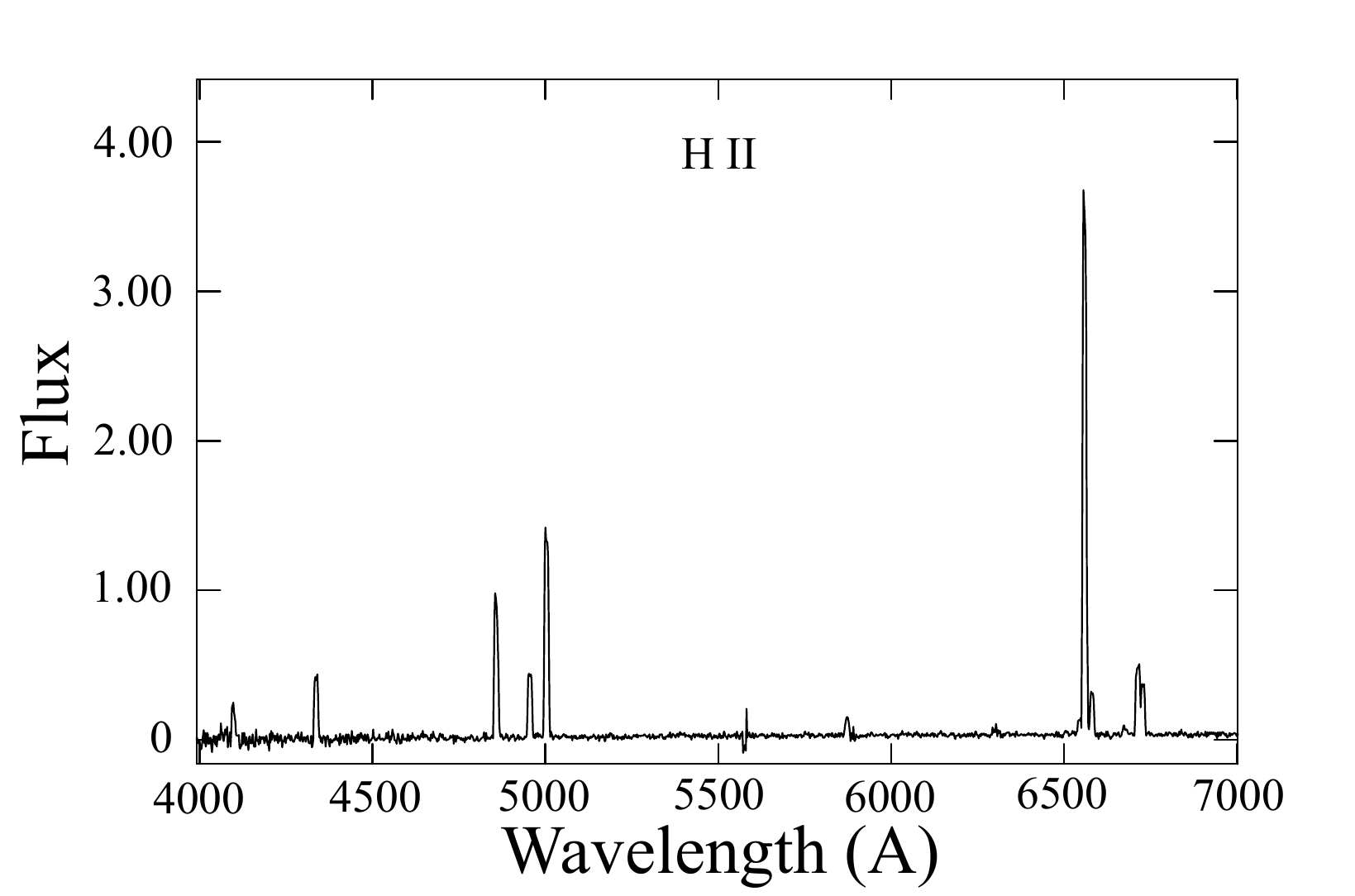}
\includegraphics[width=0.82\columnwidth]{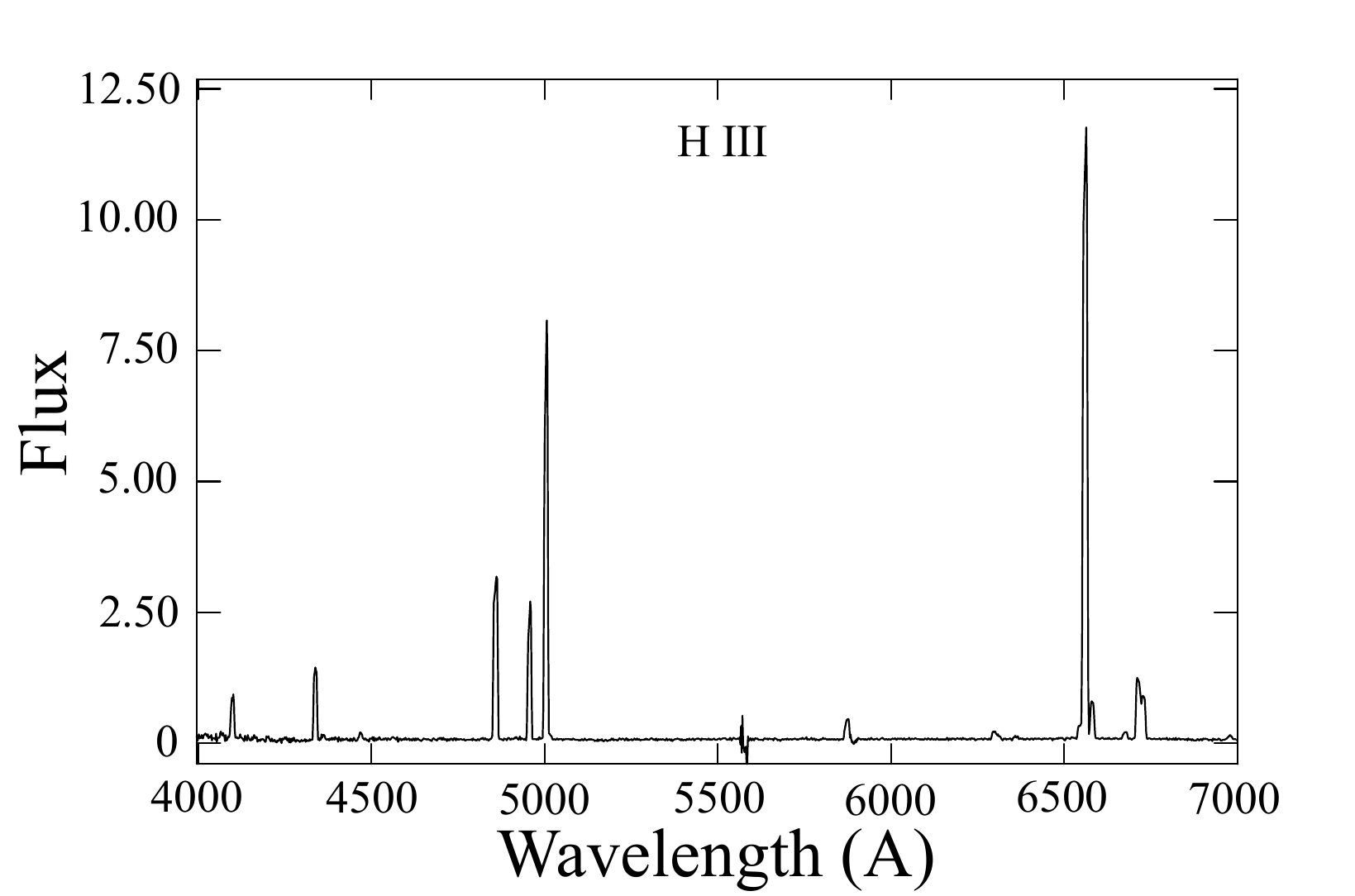}
\end{figure*}

Observed 2D spectra were bias-subtracted and flat-fielded. Afterwards the spectra were extracted, and a careful sky subtraction was made, especially in the red zone where many sky lines appear. The extraction window was chosen in such a way that all the PN nebular 
emission was included. For {\hii} regions, which are more extended, the extraction window was 3$''$ and generally it does not include the whole object.  
The standard star Ross 640 was used for flux calibration. 
Ne and Hg-Ar lamps were observed in daytime for spectral calibration. Calibrated spectra of PNe and two {\hii} regions are shown in Fig.\ref{spectra}.

Dereddened nebular line fluxes are presented in Table~\ref{intensities}, which for each object, includes the logarithmic reddening correction, c(H$\beta$), and the H$\beta$ observed flux. c(H$\beta$) was derived from the Balmer ratio H$\alpha$/H$\beta$, by assuming Case B \citep{storeyhummer95}. No underlying absorption was considered to correct H$\alpha$ and H$\beta$
emission of {\hii} regions since these lines have a very large equivalent width. In the case of PNe, the central stars are very faint, and no underlying absorption is expected  to affect the nebular lines.
An additional object observed, but not included in Table~\ref{intensities}, is PN\,26  for which only H$\alpha$ was detected with a flux of 7.93E-17  erg cm$^{-2}$ s$^{-1}$. 

Additional spectroscopic data were obtained from the literature. The line fluxes reported by \citet{hdezmartinezetal09} from VLT-FORS2 observations for PN\,4, PN\,5, PN\,6, PN\,7, PN\,10, PN\,12, PN\,14, PN\,16, PN\,18, PN\,19, PN\,21, and {\hii} 15; those by \citet{richermccall07},  from observations with the Canada-France-Hawaii (CFH) telescope and the Multi-Object Spectrograph MOS, for PN\,17 and PN\,20, and data published by \citet{peimbertetal05} for the {\hii} regions H\,V and H\,X:\ all were retrieved and physical conditions and chemical abundances recalculated, following the same procedure as for our GTC data to have a consistent data set of abundances.

\setcounter{table}{1}
\begin{table*}
\begin{tiny}
\begin{center}
\caption{Dereddened line intensities from GTC observations$^{\rm a}$.}
\label{intensities}
\begin{tabular}{lccccccccccccc}
\noalign{\smallskip} \noalign{\smallskip} \noalign{\hrule} \noalign{\smallskip}
Object& & \mc{2}{c}{PN\,2} &\mc{2}{c}{PN\,8}&\mc{2}{c}{PN\,9}&\mc{2}{c}{PN\,11}& \mc{2}{c}{PN\,13}&\mc{2}{c}{PN\,15}\\
\hline
$\lambda$&ion&I/I(H$\beta$)&err&I/I(H$\beta$)&err&I/I(H$\beta$)&err&I/I(H$\beta$)&err&I/I(H$\beta$)~&err&I/I(H$\beta$)~&err\\
\hline
3727&{\foii}&$<$ 2.40E-01&\nodata&\nodata&\nodata&\nodata&\nodata&\nodata&\nodata&noisy&\nodata&\nodata&\nodata\\
3869&{\fneiii}&3.46E-01&39&$<$ 1.00E-01&\nodata&noisy&\nodata&noisy&\nodata&\nodata&\nodata&\nodata&\nodata\\
3889&H8+{\hei}&\nodata&\nodata&2.16E-01&18&\nodata&\nodata&noisy&\nodata&noisy&\nodata&\nodata&\nodata\\
3970&H7+{\fneiii}&2.37E-01&30&1.76E-01&20&\nodata&\nodata&noisy&\nodata&noisy&\nodata&\nodata&\nodata\\
4102&H$\delta$&3.08E-01&30&2.42E-01&10&\nodata&\nodata&noisy&\nodata&2.35E-01&47&2.27E-01&30\\
4340&H$\gamma$&4.65E-01&10&4.22E-01&8&4.43E-01&30&noisy&\nodata&4.82E-01&29&4.69E-01&11\\
4363&{\foiii}&1.30E-01&40&7.28E-02&41&\nodata&\nodata&noisy&\nodata&2.44E-01&30&\nodata&\nodata\\
4471&{\hei}&1.77E-01&40& $<$ 3.00E-02&\nodata&\nodata&\nodata&noisy&\nodata&\nodata&\nodata&\nodata&\nodata\\
4686&{\heii}&$<$  6.00E-02&\nodata&7.23E-01&5&\nodata&\nodata&noisy&\nodata&$<$ 1.70E-01&\nodata&\nodata&\nodata\\
4861&H$\beta$&1.00E+00&3&1.00E+00&5&1.00E+00&5&1.00E+00&20&1.00E+00&10&1.00E+00&3 \\
4959&{\foiii}&1.43E+00&2&7.09E-01&5&\nodata&\nodata&9.92E-01&20&2.53E+00&8&\nodata&\nodata\\
5007&{\foiii}&4.49E+00&2&2.07E+00&5&2.50E-01&40&3.09E+00&3&7.25E+00&5&$<$  1.00E-02&\nodata\\
5412&{\heii}&\nodata&\nodata&5.25E-02&11&\nodata&\nodata&\nodata&\nodata&\nodata&\nodata&\nodata&\nodata\\
5755&{\fnii}&$<$ 2.00E-02&&$<$  1.90E-02&\nodata&\nodata&\nodata&4.00E-01&40&\nodata&\nodata &\nodata & \nodata\\
5876&{\hei}&1.29E-01&15&6.41E-02&25&\nodata&\nodata&noisy&\nodata&1.47E-01&14&\nodata&\nodata\\
6548&{\fnii}&3.00E-02&33&4.50E-02&22&\nodata&\nodata&3.06E+00&6&\nodata&\nodata&4.00E-02&50\\
6563&H$\alpha$&2.86E+00&2&2.85E+00&3&2.86E+00&1&2.86E+00&10&2.86E+00&5&2.86E+00&3\\
6583&{\fnii}&7.73E-02&15&1.43E-01&6&$<$  5.00E-02&\nodata&1.06E+01&1&$<$ 5.20E-02&\nodata&1.23E-01&15\\
6678&{\hei}&5.05E-02&20&2.70E-02&30&\nodata&\nodata&\nodata&\nodata&\nodata&\nodata&\nodata&\nodata\\
6716&{\fsii}&$<$ 1.50E-02&\nodata&6.60E-03&30&$<$ 4.00E-02&\nodata&4.52E-01&24&5.23E-02&19&4.83E-02&19\\
6731&{\fsii}&1.58E-02&37&1.21E-02&49&$<$  4.00E-02&\nodata&4.38E-01&25&1.77E-02&22&4.28E-02&21\\
7006&{\farv}&\nodata&\nodata&2.34E-02&38&\nodata&\nodata&\nodata&\nodata&\nodata&\nodata&\nodata&\nodata\\
7065&{\hei}&8.41E-02&20&1.66E-02&54&\nodata&\nodata&\nodata&\nodata&7.53E-02&20&\nodata&\nodata\\
7135&{\fariii}&2.34E-02&21&2.49E-02&30&\nodata&\nodata&$<$  1.90E-01&\nodata&$<$  6.60E-02&\nodata&\nodata&\nodata\\
7325&{\foii} &6.07E-02&20&2.13E-02&42&  1.00E-01&50&$<$  3.00E-01&\nodata&5.99E-02&20&\nodata&\nodata\\
\\
F(H$\beta$)$^b$& &\mc{2}{l}{5.63E-16}&\mc{2}{l}{5.94E-16}&\mc{2}{l}{2.07E-16}&\mc{2}{l}{2.50E-17}&\mc{2}{l}{1.15E-16}&\mc{2}{l}{1.12E-16}\\
c(H$\beta$)&&\mc{2}{l}{0.58}&\mc{2}{l}{0.75}&\mc{2}{l}{1.45}&\mc{2}{l}{0.126}&\mc{2}{l}{1.88}&\mc{2}{l}{1.17}\\
\hline
\end{tabular}
\end{center}
\begin{description}
\item[$^{\rm a}$] errors in \% .
\item[$^{\rm b}$] observed flux in erg cm$^{-2}$s$^{-1}$.
\end{description}
\end{tiny}
\end{table*}

\setcounter{table}{1}
\begin{table*}
\begin{tiny}
\begin{center}
\caption{Continued. Dereddened line intensities from GTC observations$^{\rm a}$.}
\label{intensities2}
\begin{tabular}{lccccccccccccc}
\noalign{\smallskip} \noalign{\smallskip} \noalign{\hrule} \noalign{\smallskip}
%\hline 
Object& &  PN\,22& &  PN\,23& &  PN\,24& &  PN\,25$^c$& &  H\,II && H\,III\\
\hline 
$\lambda$& ion&  I/I(H$\beta$)& err&  I/I(H$\beta$)& err&  I/I(H$\beta$)& err & I/I(H$\alpha$)& err & I I/I(H$\beta$)& err & I/I(H$\beta$)& err \\\hline
3727& {\foii}&\nodata&\nodata&\nodata&\nodata&\nodata&\nodata&\nodata&\nodata& 5.01E+00&10 &3.76E+00 & 7\\
3869& {\fneiii}&\nodata&\nodata&\nodata&\nodata&\nodata&\nodata&\nodata&\nodata& \nodata & \nodata & 1.29E-01& 12\\
3889& H8+{\hei}&\nodata&\nodata&\nodata&\nodata&\nodata&\nodata&\nodata&\nodata& \nodata& \nodata & 1.56E-01 & 12\\
3970& H7+{\fneiii}&\nodata&\nodata&\nodata&\nodata&\nodata&\nodata&\nodata&\nodata&\nodata& \nodata & 1.88E-01 & 10 \\
4102& H$\delta$&\nodata&\nodata&\nodata&\nodata&\nodata&\nodata&\nodata&\nodata &2.45E-01&20& 2.97E-01 & 10\\
4340& H$\gamma$&  9.80E-01& :& 2.77E-01& :&  4.60E-01& 20& \nodata&\nodata &4.93E-01 &10 & 4.83E-01 &10\\
4363& {\foiii}&\nodata &\nodata &  $<$ 2.00E-01& \nodata&  1.80E-01& 39&\nodata & \nodata   &$<$ 3.3E-02 && 2.93E-02 &20\\
4471& {\hei}&\nodata &\nodata &\nodata& \nodata&  $<$ 1.00E-01& \nodata & \nodata & \nodata&  \nodata & \nodata & 5.61E-02 & 10\\
4686& {\heii}& \nodata& \nodata&  2.93E-01& 27&  7.04E-01& 15& \nodata & \nodata &  \nodata & \nodata & \nodata &\nodata\\
4861& H$\beta$&  1.00E+00& 10&  1.00E+00& 10&  1.00E+00& 5& $<$ 2.00E-01& \nodata &   1.00E+00&\nodata&  1.00E+00& 5\\
4959& {\foiii}& \nodata& \nodata &  3.43E+00& 5&  2.51E+00& 4& \nodata & \nodata &  4.83E-01 & 7 & 7.58E-01& 5\\
5007& {\foiii}& $<$ 9.00E-02& \nodata&  1.07E+01& 4&  7.77E+00& 3& $<$ 2.00E-01& \nodata &  1.38E+00&3 & 2.28E+00 & 3 \\
5412& {\heii}& \nodata& \nodata& \nodata& \nodata &  5.00E-02& 50& \nodata & \nodata& \nodata & \nodata &\nodata & \nodata \\
5755& {\fnii}&  \nodata& \nodata& \nodata& \nodata&  2.00E-02& 50& $<$ 1.00E-01&  \nodata &\nodata & \nodata & \nodata & \nodata \\
5876& {\hei}& \nodata& \nodata &  9.87E-02& 30& $<$ 1.00E-01& \nodata& $<$ 1.00E-01& \nodata & 1.09E-01 & 10 & 1.03E-01 & 8\\
6548& {\fnii}& \nodata& \nodata &  1.50E-01& 33& 1.80E-01& 28&  1.38E+00& 10&  7.39E-02 & 12 &3.08E-02& 12\\
6563& H$\alpha$&  2.86E+00& 5&  2.86E+00& 5&  2.87E+00& 3&  1.00E+00& 10&  2.87E-01 & 3 & 2.89E+00& 3\\
6583& {\fnii}&  3.10E-01& 23&  5.32E-01& 11&  5.37E-01& 9&  4.50E+00& 3& 2.24E-01 & 10 & 1.46E-01& 8 \\
6678& {\hei}&\nodata&\nodata&\nodata&\nodata&\nodata&\nodata&\nodata&\nodata& 4.19E-2 & 20 & 3.06E-02& 12\\
6716& {\fsii}&  5.80E-01& 21&  1.67E-01& 20&  1.06E-01& 19&  1.10E-01& 20&  3.58E-01 & 10 & 2.79E-01 & 8\\
6731& {\fsii}&  5.10E-01& 20&  1.27E-01& 24&  8.69E-02& 23&  8.10E-02& 30& 2.42E-01 & 12 & 2.05E-01& 8\\
7006& {\farv}&\nodata&\nodata&\nodata&\nodata&\nodata&\nodata&\nodata&\nodata&\nodata& \nodata &\nodata & \nodata \\
7065& {\hei}& \nodata &\nodata &\nodata &\nodata &\nodata &\nodata & 4.55E-02& 44 &  9.75E-02 & 19 & 8.10E-02& 10\\
7135& {\fariii}& \nodata & \nodata&  8.82E-02& 29&  9.34E-02& 11& \nodata & \nodata &  7.92E-02 & 20 & 8.19E-02 & 10\\
7325& {\foii} &  4.00E-01&: &  9.48E-02& 32& \nodata & \nodata & \nodata & \nodata &  1.36E-01 & 25 & 9.11E-02& 10\\
\\
F(H$\beta$)$^b$& & \mc{2}{l}{8.39E-17} &  \mc{2}{l}{8.12E-17} &  \mc{2}{l}{1.45E-16} & \mc{2}{l}{9.00E-17}&  1.14E-15$^d$ & & 3.85E-15$^d$\\
c(H$\beta$)& &  \mc{2}{l}{0.61}& \mc{2}{l}{0.79} & \mc{2}{l}{0.89} & \mc{2}{l}{ $>$ 0.8}&0.39 & & 0.44 \\
\hline 
\end{tabular}
\end{center}
\begin{description}
\item[$^{\rm a}$] Errors in \%. Colons indicate errors greater than 50\%. 
\item[$^{\rm b}$] Observed {\hb} flux in erg cm$^{-2}$s$^{-1}$. 
\item[$^{\rm c}$] For PN\,25  the reported fluxes are relative to {\ha}. F({\ha}) is given.
\item[$^{\rm d}$] Extended {\hii} region. Partial {\hb} flux in a slit of 3$\times$1.5 arcsec$^2$.
\end{description}
\end{tiny}
\end{table*}

\section{The true nature of objects observed with GTC
\label{sec:PN}}

At the distance of NGC\,6822, PN candidates can be easily mixed up with very compact {\hii} regions, therefore  spectroscopic data are needed to segregate both  types of objects. In our sample of  11 PN candidates observed with GTC, we found that the {\heii} $\lambda$4686 emission line is present (and measurable) in the objects named PN\,8, PN\,23, and PN\,24, therefore their PN nature is confirmed since it is not expected that {\hii} regions emit such a high ionization line. In addition, the {\foiii} $\lambda$5007/{\hb} line ratio is larger than 3 in the objects PN\,2, PN\,11, and PN\,13, also confirming their PN nature.  The objects  PN\,9, PN\,15, PN\,22, PN\,25, and PN\,26 are very faint with observed F(H$\beta$) lower than about 2$\times10^{-16}$ erg cm$^{-2}$ s$^{-1}$, and no (or very faint) {\foiii} $\lambda$5007 or {\heii} $\lambda$4686 emission were detected, thus their PN nature is still unconfirmed, although the faintness of {\hb},  the stellar appareance, and the lack of a detectable central star support their PN nature. In particular, we want to point out the case of PN\,25, for which the {\fnii} $\lambda\lambda$6548, 6583 lines are stronger than {\ha} (see Fig.~\ref{spectra}), which makes it a good candidate for a Peimbert Type I PN. As mentioned above, only {\ha} was detected for PN\,26.

\begin{figure}
\includegraphics[width=\columnwidth]{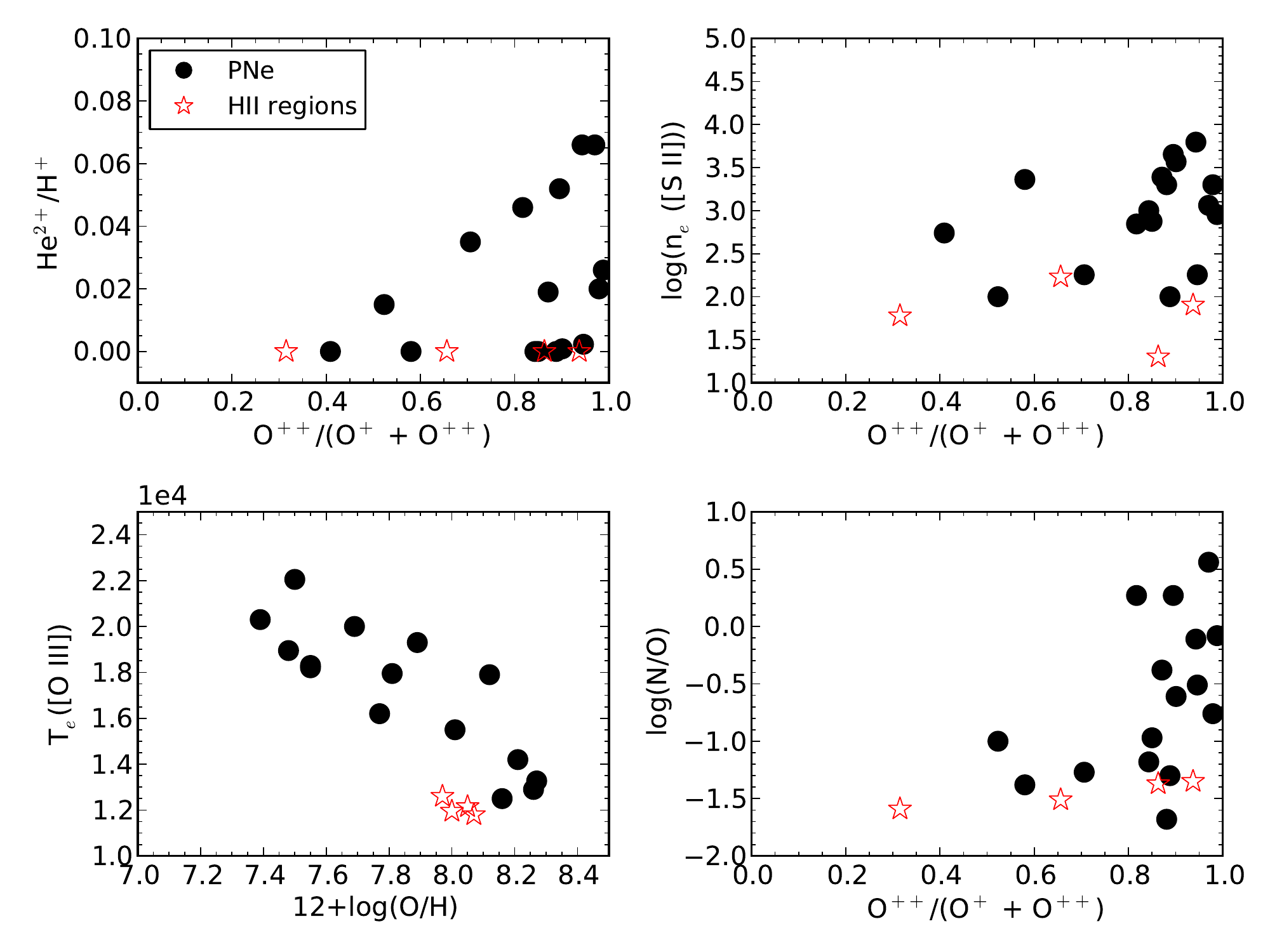}\caption{Diagrams illustrating observed properties that allow differentiatin between PNe and {\hii} regions in NGC\,6822}
\label{PN_HII}
\end{figure}

By considering the 11 objects confirmed as PNe by \citet{hdezmartinezetal09}, the two observed by \citet{richermccall07},  and the six objects observed with GTC and confirmed as PNe in this work, we find that 19 of the 26 PN candidates reported by \citet{hdezmartinezpena09} are true PNe, and the others  are highly probable PNe, thus indicating that the criteria for selecting PN candidates in external galaxies proposed by these authors are very good. However, as we see in the discussion, these criteria, which were designed to mainly select high-luminosity ({\foiii}$\lambda$5007)  PNe (useful for building the planetary nebula luminosity function, PNLF), lead to ignore a number of PNe, in particular those with low excitation and low {\foiii} $\lambda$5007 flux. Our sample therefore shows some  bias that will be discussed later. In \S~\ref{sec:totalab} we present Fig.~\ref{PN_HII}, which shows some diagrams that are useful for distinguishing between PNe and compact {\hii} regions.

\section{Physical conditions and ionic chemical abundances
\label{sec:ionic}}

All the available diagnostic line ratios were used to derive the physical conditions. Electron temperatures were calculated from {\foiii} $\lambda\lambda$4363/5007  and {\fnii} $\lambda\lambda$5755/6583 intensity ratios, while densities were derived from {\fsii} $\lambda\lambda$6717/6731 and {\fariv} $\lambda\lambda$4711/4740 intensity ratios. For H\,V and H\,X, we can also compute electron temperatures from the {\foii} $\lambda\lambda$3727/7325 intensity ratio and electron densities from the {\foii} $\lambda\lambda$3726/3729 and the {\fcliii} $\lambda\lambda$5517/5537 line ratios; the last ratio was also measured for {\hii}\,15 (see Table 2 of Hern\'andez-Mart{\'\i}nez et al. 2009). Physical conditions and ionic abundances were computed using {\pyneb}, a new python-based code for the analysis of nebular data \citep{luridianaetal15}. The atomic data set used  is listed in Table~\ref{atomic}.

In general, a two-temperature model was adopted to derive ionic abundances, where {\te}({\foiii}) was used for high ionization species (O$^{++}$, Ne$^{++}$, Ar$^{++}$, Ar$^{+3}$, He$^+$, and He$^{++}$), while {\te}{\fnii} was used for the low-ionization ones (N$^+$, O$^+$, S$^+$, and  S$^{++}$). For H\,V, the weighted average of {\te}({\fnii}) and {\te}({\foii}) was used to compute abundances of low ionization  species. When only one temperature was available, it was used for all the ions. The density derived from the {\fsii} density sensitive line ratio was used in all the cases when available. In the objects where other diagnostics were available, the densities obtained were consistent within the uncertainties with the density derived from {\fsii} lines. When no density diagnostic was observed, a density of 1000 or 2000 cm$^{-3}$ was adopted, depending on the case (see Tables~\ref{ionic_ab} and  \ref{ionic_ab_2}). 

Uncertainties in the ionic abundances were calculated by applying  Monte Carlo simulations. We generated 500 random values for each line intensity with a sigma equal to the uncertainty quoted  for the line intensities.  In Table~\ref{lines} we show the list of lines used to derive ionic abundances.

Physical conditions and ionic abundances, with their respective errors, are presented in Table~\ref{ionic_ab} for data from the VLT and the CFH telescope and for H\,V and H\,X,  and in Table~\ref{ionic_ab_2}  for data obtained with the GTC.  Although we have estimated some ionic abundances for  the {\hii} region H\,II, these values are extremely uncertain because no electron temperature could be computed for this object, therefore it will not be considered for the discussion of total abundances in {\hii} regions.

For elements heavier than He, we only considered abundances obtained from CELs. It is well known that abundances obtained from RLs are generally higher than those obtained from CELs (see \S~\ref{sec:intro}), but although some oxygen RLs have been detected in H\,V and H\,X spectra \citep{peimbertetal05}, we do not detect  oxygen RLs in the rest of our sample objects. Moreover, some authors claim that the ADF can be considered almost constant for {\hii} regions in Galactic and extragalactic domain \citep{garciarojasesteban07, penaguerreroetal12}, but this is not necessarily true for PNe \citep{mcnabbetal13}, making any conclusion obtained from assuming a ``canonical'' ADF unreliable. However, if we consider that PNe in NGC\,6822 follow the behaviour of Galactic and extragalactic {\hii} regions and the bulk of PNe in our Galaxy, the effect of considering the ADF would be much lower than the quoted uncertainties in heavy element abundance ratios and would affect elemental abundances by adding $\sim$0.2 dex to the CELs abundances.

\setcounter{table}{2}
\begin{table}
\begin{tiny}
\begin{center}
\begin{minipage}{180mm}
\caption{Atomic data set used for collisionally excited lines.}
\label{atomic}
\begin{tabular}{lcc}
 \hline
 & Transition  & Collisional \\
 Ion & probabilities & strengths \\
 \hline
N$^+$ & \citet{froesefischertachiev04} & \citet{tayal11} \\
O$^+$ & \citet{froesefischertachiev04} & \citet{kisieliusetal09} \\
O$^{2+}$ &  \citet{froesefischertachiev04} &  \citet{storeyetal14} \\
                &  \citet{storeyzeippen00} &   \\
Ne$^{2+}$ & \citet{galavisetal97} & \citet{mclaughlinbell00} \\
Ne$^{4+}$ & \citet{galavisetal97} & \citet{danceetal13} \\
          &  \citet{bhatiadoschek93} & \\
S$^+$ & \citet{podobedovaetal09} & \citet{tayalzatsarinny10} \\
S$^{2+}$ &  \citet{podobedovaetal09} & \citet{tayalgupta99} \\
Cl$^{2+}$ & \citet{mendoza83} & \citet{butlerzeippen89} \\
Ar$^{2+}$ & \citet{mendoza83} & \citet{galavisetal95} \\
          &  \citet{kaufmansugar86} & \\
Ar$^{3+}$ & \citet{mendozazeippen82} & \citet{ramsbottombell97} \\
Ar$^{4+}$ & \citet{mendozazeippen82} & \citet{galavisetal95} \\
          &  \citet{kaufmansugar86} & \\
          &  \citet{lajohnluke93} & \\
\hline
\end{tabular}
\end{minipage}
\end{center}
\end{tiny}
\end{table}

\setcounter{table}{3}
\begin{table}
\begin{tiny}
\caption{Lines used for ionic abundance determinations.}
\label{lines}
\begin{tabular}{ll}
\noalign{\hrule} \noalign{\vskip3pt}
Ion & Line \\
\noalign{\smallskip} \noalign{\hrule} \noalign{\smallskip}
N$^+$  &  {\fnii} $\lambda$$\lambda$6548, 6584 \\
O$^+$  &  {\foii} $\lambda$$\lambda$3726+29, 7320+30 \\
O$^{++}$  &  {\foiii} $\lambda$$\lambda$4959, 5007 \\
Ne$^{++}$  & {\fneiii} $\lambda$$\lambda$3868 \\
Ne$^{+4}$  &  {\fnev} $\lambda$3425   \\
S$^+$  &        {\fsii} $\lambda$$\lambda$6717+31, 4068+76 \\
S$^{++}$  & {\fsiii} $\lambda$6311      \\
Cl$^{++}$$^{\rm a}$  & {\fcliii} $\lambda$5517+37       \\
Ar$^{++}$  & {\fariii} $\lambda$$\lambda$7136, 7751     \\
Ar$^{+3}$  & {\fariv} $\lambda$$\lambda$4711+40 \\
Ar$^{+4}$ &  {\farv} $\lambda$7005          \\
\noalign{\smallskip} \noalign{\hrule} \noalign{\smallskip}
\end{tabular}
\begin{description}
\item[$^{\rm a}$] Only for VLT {\hii} region data.
\end{description}
\end{tiny}
\end{table}

\setcounter{table}{4}
\begin{table*}
\begin{tiny}
\begin{center}
\caption{Ionic abundances$^{\rm a}$ in NGC\,6822 from VLT (11 PNe and one {\hii} region) and CFH (PN\,17 and PN\,20) data and for H\,V and H\,X$^{\rm b}$.}
\label{ionic_ab}
\begin{tabular}{lcccccccc}
\noalign{\smallskip} \noalign{\smallskip} \noalign{\hrule} \noalign{\smallskip}
 Object     &        PN\,4                      &          PN\,5                        &       PN\,6                     &        PN\,7                  &        PN\,10                  &        PN\,12                 &       PN\,14  &  PN\,16  \\
\noalign{\smallskip} \noalign{\hrule} \noalign{\smallskip}
{\te}({\foiii}) (K)  & 17950$\pm$1250           &  $<$24600                             & 12500$\pm$750           &  13270$\pm$2400       & 18200$\pm$1150          & 12900$\pm$650    & 17900$\pm$1000  & 14200$\pm$700            \\
{\te}({\fnii}) (K)   &  14200$\pm$2650    &   15350$\pm$1500           & 11600$\pm$3100             & \nodata                    & \nodata                                        & \nodata          &  20500$\pm$2500    &  \nodata                          \\
{\elecd}({\fsii}) (cm$^{-3}$)  &  2450$^{+1800}_{-1000}$ & 700$^{+1100}_{-450}$ & 3700$^{+5200}_{-2150}$& 2300$^{+7000}_{-1700}$& 1000 (adopted)  &  2000 (adopted)      &  4500$^{+8100}_{-2900}$     & 900$^{+1650}_{-600}$   \\
{\elecd}({\fariv}) (cm$^{-3}$)  & 1900:    &    \nodata           &  \nodata           &  \nodata            & \nodata           &  1700$^{+4600}_{-1250}$           &  11500:   & 4500$^{+7500}_{-2800}$      \\
\noalign{\smallskip} \noalign{\hrule} \noalign{\smallskip}
Ion & \multicolumn{8}{c}{12+log (X$^{+i}$/H$^+$)} \\
\noalign{\smallskip} \noalign{\hrule} \noalign{\smallskip}
             He$^+$     & 10.81$\pm$0.04                & 10.80$\pm$0.16                        & 10.98$\pm$0.04          & 11.09$\pm$0.10        & 11.15$\pm$0.04        & 10.91$\pm$0.06          & 10.72$\pm$0.07& 11.03$\pm$0.05                \\
         He$^{++}$      & 10.29$\pm$0.04                & 10.66$\pm$0.05        & 8.93$^{+0.16}_{-0.25}$  & $<$9.93                       & $<$9.36                       & 10.30$\pm$0.06  & 10.72$\pm$0.02& 10.42$\pm$0.05                \\
             N$^+$      & 6.52$^{+0.18}_{-0.15}$        & 7.40$\pm$0.09 & 6.39$^{+0.32}_{-0.25}$ & 6.48$^{+0.22}_{-0.13}$& 5.46$\pm$0.12& 5.81$\pm$0.09           & 7.31$^{+0.07}_{-0.04}$& 6.21$\pm$0.05           \\
             O$^+$      &  6.85$^{+0.27}_{-0.24}$       & 7.05$\pm$0.24 & 7.15:                   & 7.87:                 & 6.74$\pm$0.13 & 6.53$^{+0.24}_{-0.19}$        & 6.94$^{+0.11}_{-0.08}$&  6.24$\pm$0.13          \\
         O$^{++}$       &7.68$\pm$0.07          & 7.70$\pm$0.11         & 8.11$\pm$0.07   & 8.01$^{+0.23}_{-0.17}$& 7.47$\pm$0.06 & 8.20$\pm$0.08         & 7.87$\pm$0.05&8.15$\pm$0.05             \\
        Ne$^{++}$       &6.93$\pm$0.08          & \nodata                       & 7.34$\pm$0.08   & \nodata                       & 6.65$\pm$0.08 & 7.36$\pm$0.09         & \nodata &7.41$\pm$0.07  \\
        Ne$^{+4}$       & 6.52$\pm$0.16                 &  \nodata                      & \nodata                 & \nodata                       & \nodata                       & \nodata                         & \nodata        & \nodata      \\
              S$^+$     &5.20$^{+0.19}_{-0.16}$         & 5.79$\pm$0.10         & 5.16:                   & 4.85$^{+0.28}_{-0.24}$& \nodata               & 4.38$^{+0.26}_{-0.23}$  & 4.86$^{+0.25}_{-0.16}$ &4.67$^{+0.11}_{-0.09}$        \\
         S$^{++}$       &5.89$\pm$0.10          & 6.21$^{+0.20}_{-0.25}$&  5.91$^{+0.16}_{-0.21}$ &5.88$\pm$0.29& \nodata                 & 6.15$\pm$0.12         &5.56$^{+0.16}_{-0.25}$ &5.86$^{+0.12}_{-0.15}$ \\
        Ar$^{++}$       &5.31$\pm$0.07          & 5.64$\pm$0.11 & 5.55$\pm$0.06 & 5.79$\pm$0.19   & 4.88$^{+0.12}_{-0.15}$& 5.62$\pm$0.09         & 5.39$\pm$0.09 &5.41$\pm$0.08                 \\
        Ar$^{+3}$       &5.22$\pm$0.09          &  \nodata                      & 5.23$\pm$0.17   & \nodata                       & \nodata                       & 5.63$\pm$0.11           & 5.32$\pm$0.09 &5.67$\pm$0.08          \\
        Ar$^{+4}$       & 4.50$^{+0.13}_{-0.17}$        & \nodata               & \nodata                 & \nodata                       &  \nodata                      & \nodata                         & 5.13$^{+0.10}_{-0.13}$ & \nodata      \\
\hline
\noalign{\smallskip} \noalign{\hrule} \noalign{\smallskip}
  Object             &      PN\,18              &     PN\,19                    &         PN\,21          &        {\hii}\,15     &          PN\,17 &    PN\,20    &  H\,V   & H\,X    \\
\noalign{\smallskip} \noalign{\hrule} \noalign{\smallskip}
{\te}({\foiii}) (K)  & 18950$\pm$1900    &  22050$\pm$3300      &  16200$\pm$1200   &  12150$\pm$200              &   19300$\pm$3000         & 15500$\pm$4000    & 11800$\pm$250  & 11950$\pm$470      \\
{\te}({\foii}) (K)   &  \nodata                 &  \nodata           & \nodata                                 &  \nodata              & \nodata                & \nodata    &  15300$\pm$900 & 14300$\pm$1100     \\
{\te}({\fnii}) (K)   &  \nodata                 &  \nodata           & \nodata                                 &  12000$\pm$1800               & \nodata                 & \nodata    & 15800$\pm$2850  & \nodata    \\
{\elecd}({\fsii}) (cm$^{-3}$)  & 1150$^{+3400}_{-850}$& $<$100   &  180$^{+470}_{-150}$ &  170$^{+240}_{-100}$          & 180$^{+370}_{-120}$           & 2000 (adopted)   &  80$^{+110}_{-50}$   &  $<$20    \\
{\elecd}({\foii}) (cm$^{-3}$)  & \nodata   &  \nodata        &  \nodata            &  \nodata          &  \nodata           &  \nodata    & 70$^{+70}_{-40}$  & 50$^{+70}_{-30}$    \\
{\elecd}({\fcliii}) (cm$^{-3}$) & \nodata   &  \nodata           &  \nodata                         &  300$^{+800}_{-210}$          &  \nodata           &  \nodata     & 600$^{+1500}_{-430}$  & $<$20   \\
{\elecd}({\fariv}) (cm$^{-3}$)  & \nodata   &  \nodata        &  \nodata            &  \nodata          &  \nodata           &  \nodata    & \nodata  &  \nodata   \\
\noalign{\smallskip} \noalign{\hrule} \noalign{\smallskip}
             He$^+$     & $<$10.76                      & 11.09$^{+0.15}_{-0.26}$       & 11.11$^{+0.06}_{-0.09}$& 10.90$\pm$0.02 & 10.98$^{+0.11}_{-0.16}$ & 10.91$^{+0.05}_{-0.07}$   &  10.92$\pm$0.02   & 10.90$\pm$0.04        \\
         He$^{++}$      & 10.82$\pm$0.02                & 10.18$\pm$0.07                & 9.37$^{+0.16}_{-0.26}$& \nodata                 & 10.54$\pm$0.07        &9.97$^{+0.17}_{-0.30}$   &  \nodata   & \nodata         \\
             N$^+$      & 6.40$\pm$0.09         & 6.15$^{+0.14}_{-0.10}$         & 5.88$^{+0.08}_{-0.10}$& 6.08$^{+0.16}_{-0.14}$& 6.06$\pm$0.15 & 5.38$^{+0.17}_{-0.29}$   &  5.57$\pm$0.15   & 5.80$\pm$0.08        \\
             O$^+$      & 5.72$^{+0.18}_{-0.14}$        & 7.15$^{+0.21}_{-0.24}$        & 6.50$\pm$0.16   & 7.58$^{+0.27}_{-0.21}$        & 7.28$^{+0.18}_{-0.19}$ & 7.06$^{+0.12}_{-0.17}$   & 6.87$^{+0.25}_{-0.20}$    & 7.14$\pm$0.13        \\
         O$^{++}$       & 7.23$^{+0.10}_{-0.08}$        & 7.19$^{+0.17}_{-0.09}$        & 7.74$\pm$0.08   & 7.86$\pm$0.03         & 7.66$^{+0.18}_{-0.15}$        & 7.93$\pm$0.02   & 8.04$\pm$0.03    & 7.94$\pm$0.06     \\
        Ne$^{++}$       & 6.65$^{+0.12}_{-0.14}$        & \nodata                               & 6.96$^{+0.13}_{-0.10}$& 7.12$\pm$0.03   &       6.99$^{+0.26}_{-0.21}$ & 7.40$^{+0.10}_{-0.13}$   & 7.30$\pm$0.04    & 7.18$\pm$0.07             \\
              S$^+$     & 5.09$^{+0.21}_{-0.17}$ & 5.76:                                & 5.06$\pm$0.09   & 5.45$^{+0.15}_{-0.13}$        & 5.78$^{+0.13}_{-0.11}$ & 4.83$^{+0.08}_{-0.10}$   &  5.09$^{+0.15}_{-0.12}$   &  5.29$^{+0.11}_{-0.08}$    \\
         S$^{++}$       &\nodata                                &  5.82:                                & $<$3.61                         & 6.30$\pm$0.06         & \nodata               & \nodata   &  6.34$\pm$0.05   & 6.35$\pm$0.10            \\
        Cl$^{++}$       & \nodata                       & \nodata                               & \nodata                 &  4.40$^{+0.07}_{-0.09}$& \nodata              & \nodata   & 4.43$\pm$0.06    & 4.38$^{+0.11}_{-0.15}$           \\
        Ar$^{++}$       & \nodata                               & 5.30$^{+0.17}_{-0.15}$        & 5.40$\pm$0.11   & 5.76$\pm$0.04         & \nodata               & 5.63$\pm$0.06   &  5.76$\pm$0.03   &  5.75$\pm$0.05           \\
        Ar$^{+3}$       &  \nodata                              & \nodata                               & \nodata                 & \nodata                               & \nodata               & \nodata   &  4.80$\pm$0.10   &  \nodata         \\
\noalign{\smallskip} \noalign{\hrule} \noalign{\smallskip}
\end{tabular}
\end{center}
\begin{description}
\item[$^{\rm a}$] Abundances in 12+log (X$^{+i}$/H$^+$).  H\,V and H\,X line intensities were adopted  from \citet{peimbertetal05}.
\item[$^{\rm b}$] Colons indicate errors larger than a factor of two. 
\end{description}
\end{tiny}
\end{table*}

\setcounter{table}{5}
\begin{table*}
\begin{tiny}
\begin{center}
\caption{Ionic abundances$^{\rm a}$ in NGC\,6822 objects from GTC data$^{\rm b}$.}
\label{ionic_ab_2}
\begin{tabular}{lcccccccc}
\noalign{\smallskip} \noalign{\smallskip} \noalign{\hrule} \noalign{\smallskip}
Object      &        PN\,2 &          PN\,8 &       PN\,11 &        PN\,13 &        PN\,23&        PN\,24 &       H\,II & H\,III\\
\noalign{\smallskip} \noalign{\hrule} \noalign{\smallskip}
{\te}({\foiii}) (K)  &  18300$\pm$3800  & 20300$\pm$3900        &   \nodata                 &   20000$\pm$3000      &  $<$14700             &  16300$\pm$3300                 &  $<$16700  & 12600$\pm$1000       \\
{\te}({\fnii}) (K)   &   \nodata                        &  \nodata               & 17650$\pm$5750        &   \nodata             &  \nodata                       &  15850$\pm$8700               &  \nodata   &  \nodata      \\
{\elecd}({\fsii}) (cm$^{-3}$)  &  750$^{+3000}_{-600}$ & 6250: & 550$^{+1250}_{-400}$&   $<$100                & 100$^{+300}_{-75}$    &  200$^{+550}_{-140}$  &  $<$100   & 60$^{+100}_{-40}$      \\
\noalign{\smallskip} \noalign{\hrule} \noalign{\smallskip}
Ion & \multicolumn{7}{c}{12+log (X$^{+i}$/H$^+$)} \\
\noalign{\smallskip} \noalign{\hrule} \noalign{\smallskip}
              He$^+$    & 11.26$\pm$0.13                & 10.57$^{+0.13}_{-0.11}$       & \nodata                         & 11.08$\pm$0.06        & 10.89$^{+0.13}_{-0.16}$       & $<$11.15        & 11.02$\pm$0.06        & 10.96$\pm$0.02\\
         He$^{++}$      & $<$9.73                               & 10.82$\pm$0.04                & \nodata                         & $<$10.19              & $>$10.40      & 10.79$^{+0.06}_{-0.08}$& \nodata& \nodata       \\
             N$^+$      & 5.68$^{+0.22}_{-0.14}$        & 5.84$^{+0.18}_{-0.15}$        & 7.78$^{+0.21}_{-0.18}$  & 5.41$\pm$0.20                 & $>$6.62       & 6.60:           & $>$6.18 & 6.15$^{+0.10}_{-0.08}$      \\
             O$^+$      &  6.72:                                & 5.83:                         & $<$7.51                         & 6.71$^{+0.32}_{-0.18}$&  $>$7.43                              & \nodata         & $>$7.49 & 7.74$^{+0.14}_{-0.11}$      \\
         O$^{++}$       & 7.47$^{+0.21}_{-0.15}$        & 7.05$^{+0.26}_{-0.15}$         & 7.35$^{+0.31}_{-0.24}$        & 7.61$^{+0.19}_{-0.10}$&  $>$8.07      & 7.83$^{+0.29}_{-0.17}$& $>$7.07 & 7.59$^{+0.12}_{-0.08}$\\
        Ne$^{++}$       & 6.74$^{+0.33}_{-0.30}$        & $<$6.08                               & \nodata                         & \nodata                       & \nodata                               & \nodata                 & \nodata       & 6.77$^{+0.15}_{-0.11}$        \\
              S$^+$     & 4.44$^{+0.25}_{-0.17}$        & 4.45$^{+0.38}_{-0.35}$         & 5.90$\pm$0.23         & 4.78$^{+0.14}_{-0.11}$&  $>$5.51      & 5.28$^{+0.31}_{-0.23}$& $>$5.72 &  5.85$^{+0.09}_{-0.07}$\\
         S$^{++}$       & \nodata                               & \nodata                               &  \nodata                                & \nodata                       & \nodata                         & \nodata                       & \nodata       & \nodata \\
        Ar$^{++}$       & 4.83$^{+0.18}_{-0.16}$        & 4.79$^{+0.22}_{-0.19}$        & 5.77$\pm$0.28           & $<$5.22                       &  $>$5.56      & 5.51$^{+0.20}_{-0.14}$& $>$5.43 & 5.65$^{+0.10}_{-0.07}$\\  
        Ar$^{+4}$       & \nodata                               & 4.98$^{+0.23}_{-0.26}$        & \nodata                         & \nodata                       & \nodata                               & \nodata                 & \nodata       & \nodata       \\
\noalign{\smallskip} \noalign{\hrule} \noalign{\smallskip}
\end{tabular}
\end{center}
\begin{description}
\item[$^{\rm a}$] Abundances in 12+log (X$^{+i}$/H$^+$).
\item[$^{\rm b}$] Colons indicate errors larger than a factor of two.
\end{description}
\end{tiny}
\end{table*}

\section{Total abundances and AGB nucleosynthesis 
\label{sec:totalab}}

To derive the total abundance of an element, we must add the determined ionic abundances  and correct for the unseeing ions by employing ionization correction factors (ICF). Very recently, a new set of ICFs for PNe has been presented by \citet{delgadoingladaetal14}. These ICFs were computed from a large grid of photoionization models, covering a wide range of values in the parameter space, thus improving significantly the previous ICF sets for PNe in the literature.  We checked the validity range of the ICFs, depending on the excitation of each PN. In the majority of the cases, the ICFs can be applied; however, four PNe (PN\,12, PN\,16, PN\,18, and PN\,21) showed O$^{2+}$/(O$^+$ + O$^{2+}$) $\geq$ 0.95, which is out of the validity range of the ICF proposed by \citet{delgadoingladaetal14} for N, S, and Ar. In such cases we compared the results obtained with the ICF provided by \citet{delgadoingladaetal14} for N with the classical ICF scheme, i.e., N/O$\sim$N$^{+}$/O$^+$, and we found average differences of about $\pm$0.10 dex, which are within the uncertainties of the total abundance determinations; moreover, we do not find any unexpected behaviour of the abundances of Ar and S on these objects (see \S~\ref{alpha}), and therefore, to have a homogeneous analysis, we  applied the \citet{delgadoingladaetal14} scheme to these objects. 
For {\hii} regions we used the ICF expressions for mid-metallicity (7.6$<$(12+log (O/H) $<$8.2) proposed by \citet{izotovetal06}. The intrinsic uncertainties of the \citet{delgadoingladaetal14} ionization correction factors are not propagated in our analysis and only error propagation of the ionic abundances was considered in our Monte Carlo simulations.

Total abundances for 19 PNe are listed in Table~\ref{total_ab} for data from the VLT and CFH observations and in Table~\ref{total_ab_2} for data from  the GTC. In addition, the chemical abundances of four {\hii} regions -- {\hii}\,15 and H\,III observed by us and H\,V and H\,X from the work by \citet{peimbertetal05} for which we recalculated the abundances -- have been included for comparison. The behaviour of elemental abundances are shown in Figs.~\ref{L-O},~\ref{N-O}, \ref{Ne-O}, ~\ref{Ar-O}, and \ref{S-O}, and they are discussed in the following.  In these plots we have not included uncertain determinations of abundances (those with errors quoted as :) or the upper and lower limits to the abundances.

In Tables~\ref{total_ab} and ~\ref{total_ab_2}  we also list the total luminosity L(H$\beta$) of PNe in solar units for all the objects for which F(H$\beta$) and c(H$\beta$)  are published. A distance of 459.19 kpc was assumed for NGC\,6822 \citep{gierenetal06}. After determining physical conditions and chemical abundances, in Fig.~\ref{PN_HII} we present some diagrams comparing the properties of PNe and {\hii} regions that clearly show the differences between these two type of photoionized nebulae. It is evident that spectroscopic analysis allows a reliable classification on a very solid basis.

\subsection{O abundances}

\begin{figure}
\includegraphics[width=\columnwidth]{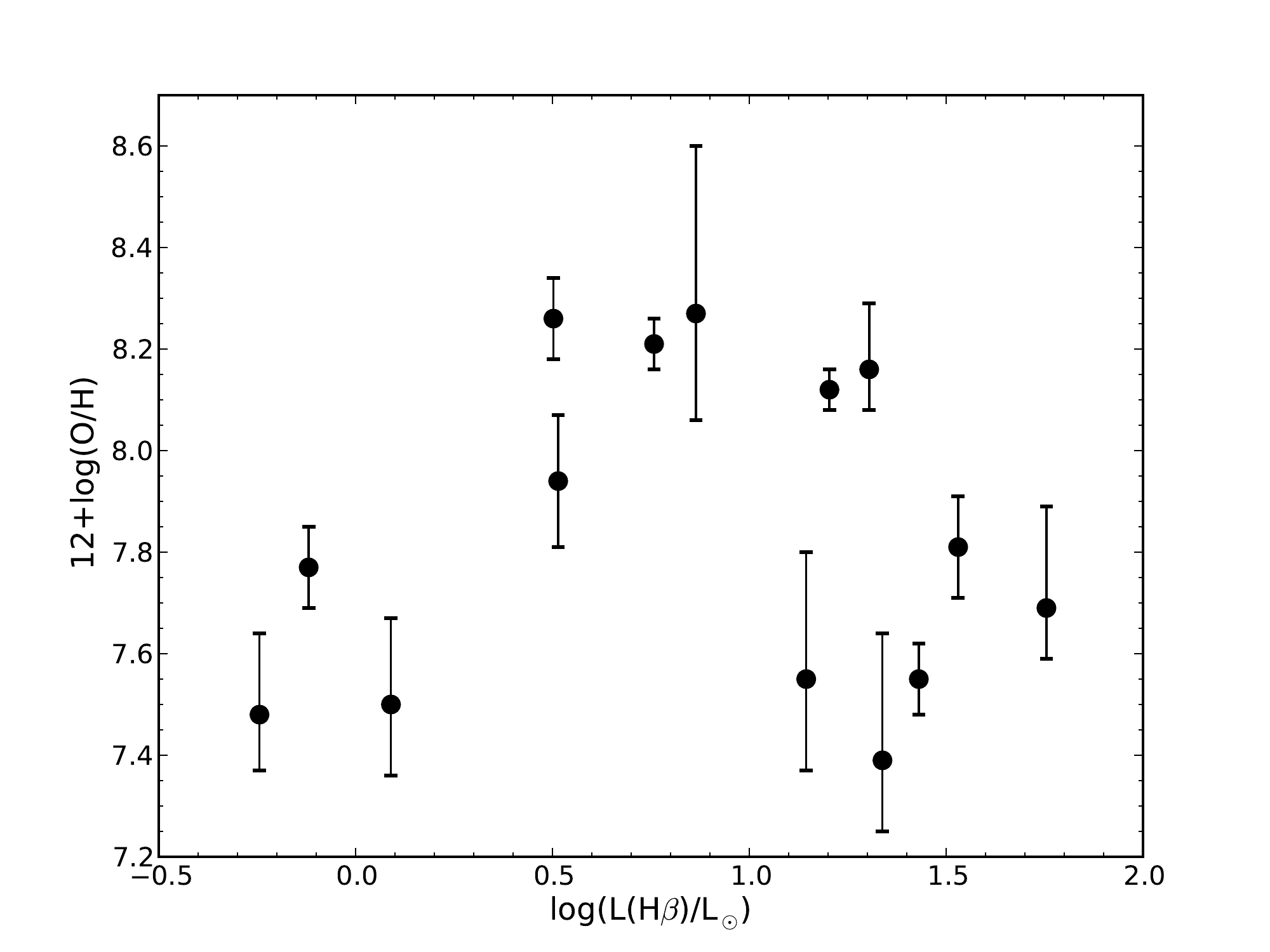}
\includegraphics[width=\columnwidth]{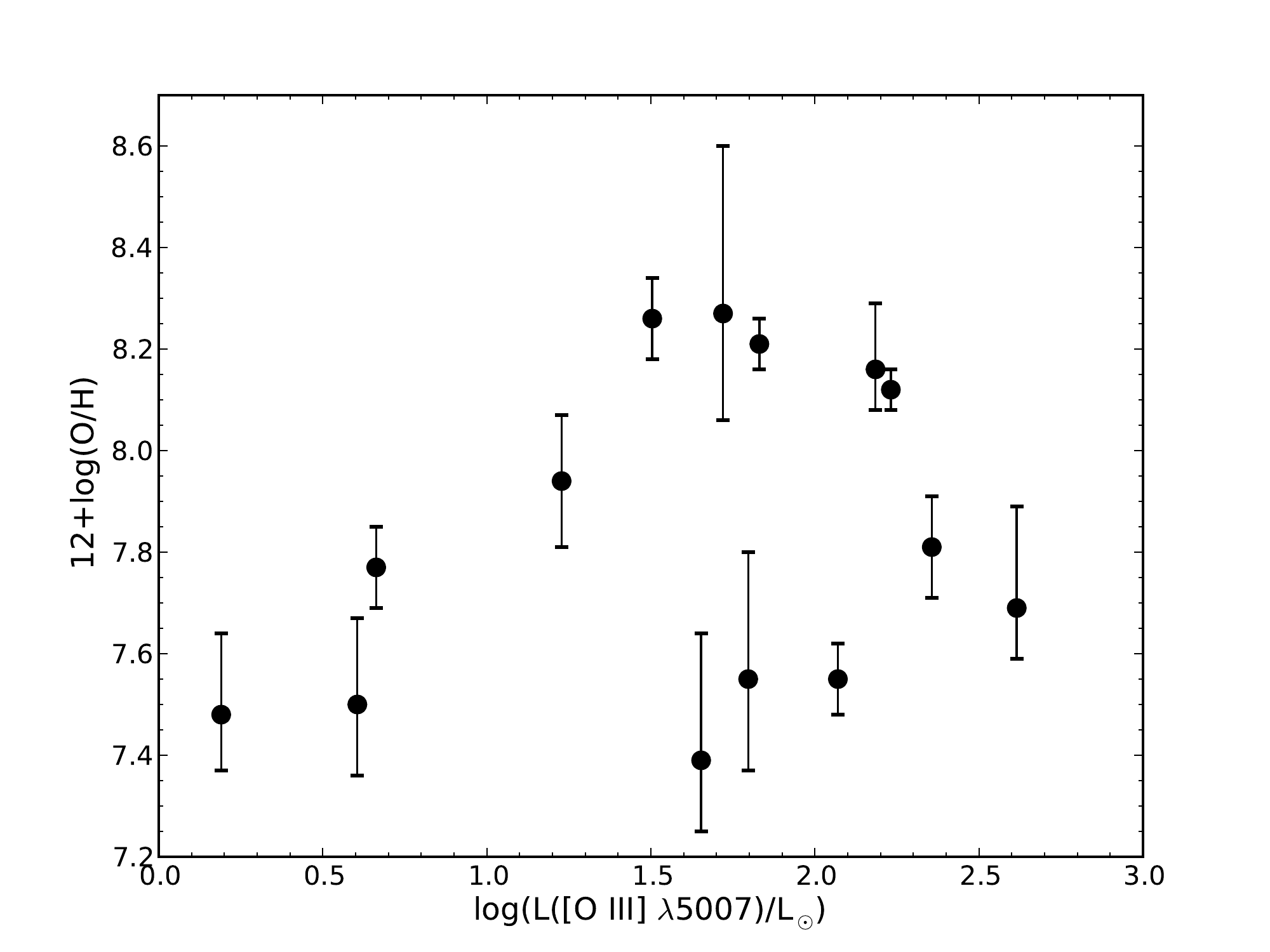}
\caption{Above: 12+log (O/H) vs. L(H$\beta$); below: 12+log (O/H) vs. L({\foiii}5007) for PNe in NGC\,6822.}
\label{L-O}
\end{figure}

\begin{figure}
\includegraphics[width=\columnwidth]{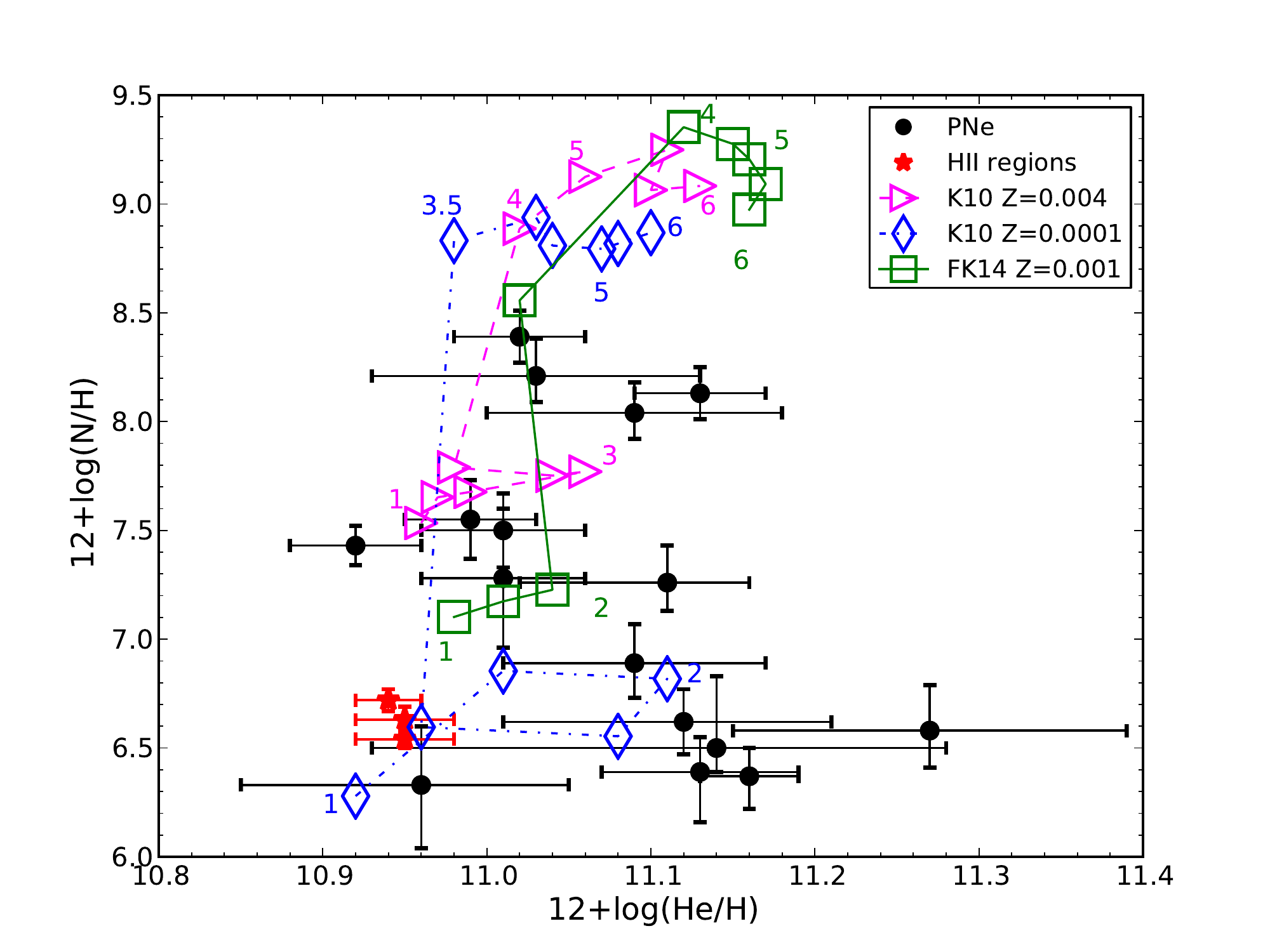}
\includegraphics[width=\columnwidth]{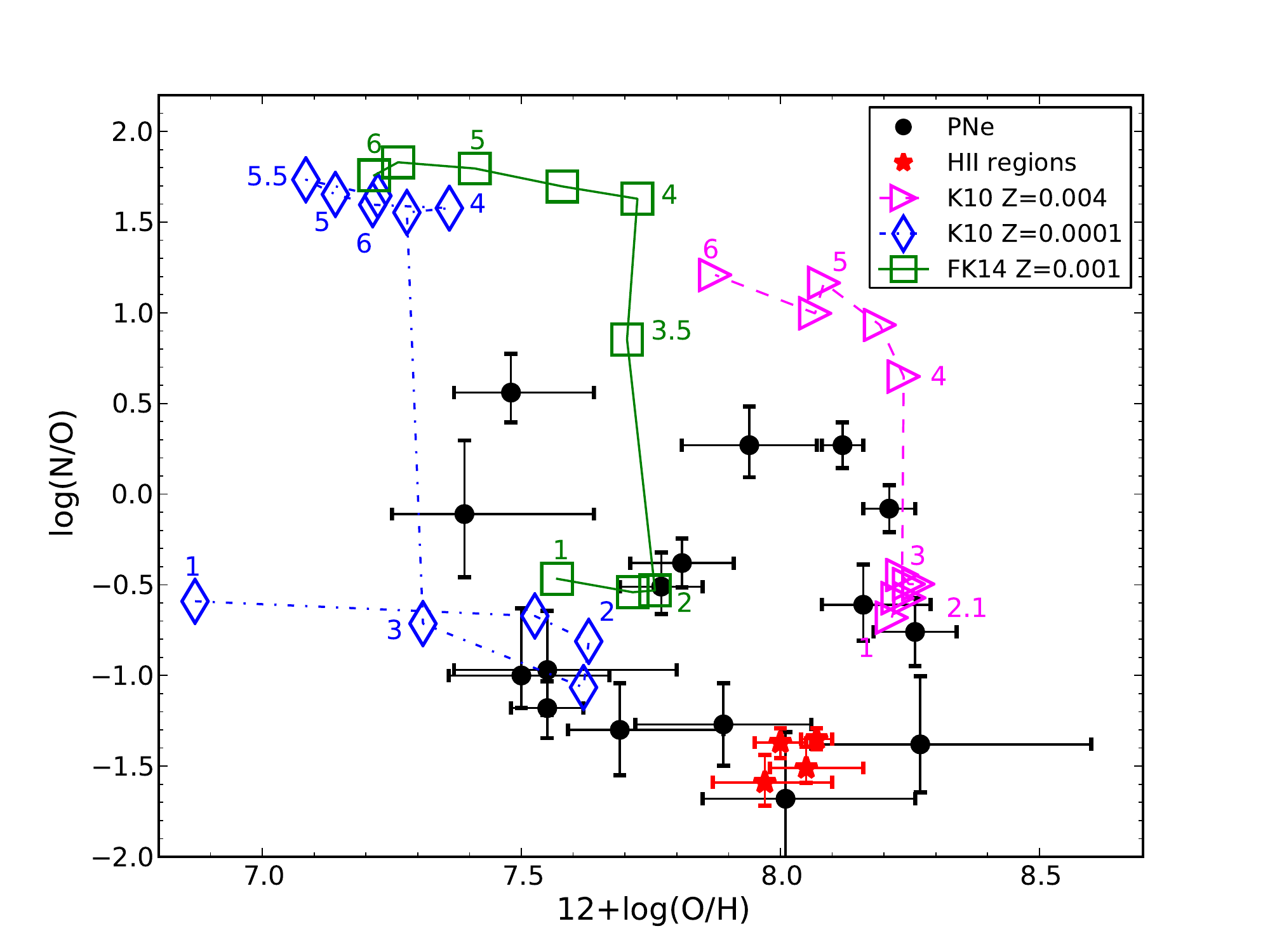}
\caption{ 12+log (N/H) vs. 12+log (He/H) (upper panel) and log (N/O) vs. 12+log (O/H)  (lower panel) for PNe (in black) in NGC\,6822. In red the {\hii} regions. Stellar evolution models for different metallicities by \citet[][K10]{karakas10} (Z=0.0001, blue diamonds; Z=0.004, magenta triangles) and \citet[][FK14]{fishlocketal14} (Z=0.001, green squares) have been included. Masses of the progenitors stars are shown for some models. See text for discussion. }
\label{N-O}
\end{figure}

In the upper panel of Fig.~\ref{L-O}, we present 12+log (O/H) vs. log (L(H$\beta$)/L$_\odot$). In Tables~\ref{total_ab} and~\ref{total_ab_2}, it is observed that PNe in NGC\,6822 show L(H$\beta$) smaller than 60 L$_\odot$, while {\hii} regions (even the compact ones) are expected to be much brighter; for instance, the compact {\hii}\,15 has L(H$\beta$) larger than 300 L$_\odot$. L(H$\beta$) for the most extended {\hii} regions are not presented because only a fraction of the nebula was observed. Regarding O abundances, 12+log (O/H) values in PNe show a wide range from 7.39 to 8.27, with the two brightest PNe showing intermedium O abundances. We do not find a particular trend between L(H$\beta$) and the O/H abundance, although it is worth noticing that the faintest objetcs show very low O abundances. The same is observed in the lower panel of Fig.~\ref{L-O}, where  we plot L({\foiii}  $\lambda$5007) instead of L(H$\beta$) and where no particular trend is
again found. 

In their Fig. 4,  \citet{corradietal15b} found for M\,31 PNe that there seems to be a tendency toward decreasing metallicity with decreasing {\foiii} luminosity. They also seem to find a similar tendency when plotting M\,33 data from \citet{magrinietal09}. Similar results have already been indicated by \citet{stasinskaetal98} for the LMC, SMC, and the Milky Way. From our data, this tendency is not evident, although as pointed out above,  the faintest PNe of our sample  show  low metallicities, while the bright ones have a wide spread in O/H abundances. 

\subsection{He, N, and O}

Figure~\ref{N-O} shows the behaviour of 12+log (N/H) as a function of 12+log (He/H) (upper panel) and log (N/O) ratio vs. 12+log (O/H) (lower panel) for PNe (black dots) and {\hii} regions (red stars). No correlation between these quantities is evident. In comparison with the {\hii} regions, most of the PNe are both He-rich and N-rich, as occurs in all the samples analysed by many authors for different galaxies. This indicates that central stars of PNe are providing large amounts of He and N to the ISM as a consequence of the first and second dredge-up events experienced  during the AGB phase. The second dredge-up occurs for the more massive stars with M $>$ 3 M$_\odot$. Such dredge-up processes enrich  the stellar surface with N produced  via the CNO cycle, but it is apparent from this figure that, since no correlation exists, the ON-cycle is not a significant contributor to the N nucleosynthesis yield, and most of the N should be contributed by the CN-cycle in these stars.  However, a note of caution should be introduced here, because \citet[][and references therein]{karakasetal09} point out that PNe with rapid rotating progenitor stars can also reflect high He and N abundances in the gas owing to rotation changes the internal structure of the star, resulting in high He/H and N/O ratios in the surface, before the AGB phase. 
Extreme N enrichment is observed in some PNe (they would be Peimbert Type I PNe)  whose central stars (in principle the most massive ones)  produce extra N enrichment  by experiencing envelope-burning conversion to N of dredged-up primary C (phenomenon known as "hot-bottom burning", hereinafter HBB). Additionally, the role of the binary evolution of low-mass stars cannot be ruled out as the origin of a fraction of highly N-rich PNe \citep[see e.~g.][and references therein]{moedemarco06}. If we consider Type I PNe as those with a N/O abundance ratio greater than 0.5 \citep[see discussion in][for adopting this value]{hdezmartinezetal09}, we found that there are six Type I PNe in our sample,  representing one third of the whole analysed sample, although two of them have very uncertain N and/or O abundances and are not represented in Fig.~\ref{N-O}. Interestingly, in the lower panel of this figure, it is evident that these Type I PNe occur at any value of O/H (that is, at any metallicity), and they are not restricted to the young (supposedly more massive) high-metallicity objects. 

As said above, to check the influence of the selected ICF scheme on our results, we also computed total N abundances using the classical ICF scheme of N/O$\sim$N$^+$/O$^+$, and we did not find any significant differences (on average, less than $\pm$0.1 dex) between using the \citet{delgadoingladaetal14} scheme and the classical one, with the exception of PN\,11, whose N/O ratio decreases $\sim$0.26 dex using the classical ICF scheme, but this object is not included in Fig.~\ref{N-O} because it shows a very uncertain O/H ratio.

In Fig.~\ref{N-O} we have also included the behaviour predicted by the stellar evolution models by \citet{karakas10} and \citet{fishlocketal14} for different metallicities. Model values correspond to the surface stellar abundances at the end of the AGB phase. The initial masses of stars have been tagged. In the upper panel of this figure, it is apparent that our objects have larger He/H abundances than what is predicted by the models. Only models with metallicity $Z$=0.0001 would be in agreement with the He abundance shown by PNe with 12+log(N/H) lower than 7.0. 

In the lower panel of Fig.~\ref{N-O}, it is found that our objects appear well demarcated by models with $Z$ between 0.0001 and 0.004.  As expected, models predict that more massive stars produce  larger amounts of N, similar to the values shown by Type I PNe. According to these models, our objects would have had initial masses lower than 4 M$_\odot$; however, it should be noticed that no model predicts the low N/O ratio (log (N/O) $>$ -1.0) shown by several of our PNe. Such stars should have had initial masses lower than 1 M$_\odot$,  or the stellar evolution models could be predicting yields for N that are too
large. 
The behaviour of these models is discussed further in \S~5.4.

\subsection{The alpha elements: Ne, Ar, and S
\label{alpha}}

\begin{figure}
\includegraphics[width=\columnwidth]{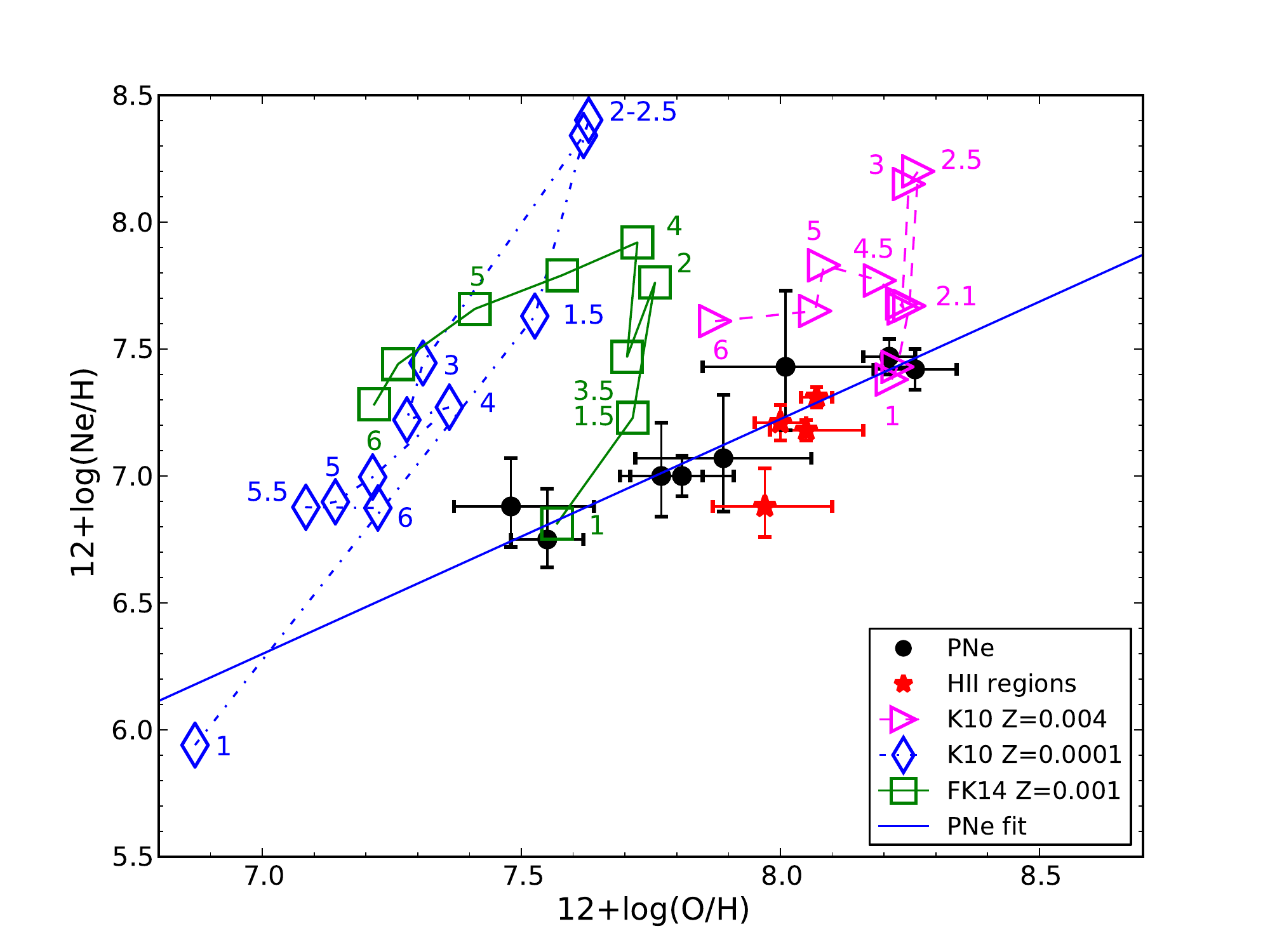}
\caption{12+log (Ne/H) vs. 12+log (O/H), for PNe (in black) in NGC\,6822. In red the {\hii} regions. The linear fit (including only PNe data) corresponds to 12+log (Ne/H)=(0.925$\pm$0.071)$\times$(12+log (O/H)) $-$ (0.176$\pm$0.566), with r= 0.88. Stellar evolution models for different metallicities, by \citet[][K10]{karakas10} (Z=0.0001, blue diamonds; Z=0.004, magenta triangles) and \citet[][F14]{fishlocketal14} (Z=0.0001, green squares) have been included. See text for discussion.}
\label{Ne-O}
\end{figure}

 The Ne/H vs. O/H abundance behaviour is shown in  Fig.~\ref{Ne-O}. Values derived for PNe and {\hii} regions are presented. A tight correlation is found between Ne and O abundances as occurs in any other sample of PNe and {\hii} regions \citep{henry89, garciahdezgorny14}. This has been interpreted as Ne not being strongly altered by AGB evolution so that O and Ne grow in lock-step during the chemical evolution of galaxies. In the case of our PNe sample, the slope of the correlation is 0.93$\pm$0.07, which is similar to  the slopes found for other PN samples; for instance, \citet{leisydennefeld06}  report slopes of 1.13 for PNe in the LMC and 1.01 in the SMC. 

Similar to what has been done with N, we checked the Ne/H ratio obtained using the ICF scheme by \citet{kingsburghbarlow94}. The differences between this ICF scheme and the one assumed in this work \citep{delgadoingladaetal14} are, on average, much smaller than the derived uncertainties ($<$0.05 dex) with the exception of two objects, PN\,17 and PN\,6, for which the Kingsburgh \& Barlow scheme gives Ne/H ratios that are 0.48 and 0.26 dex larger, respectively.  PN\,6 is not shown in Fig.~\ref{Ne-O} owing to its uncertain Ne/H ratio.

In Fig.~\ref{Ne-O}  we have also included the predictions of stellar evolution models by \citet{karakas10} and \citet{fishlocketal14} for different metallicities. These models consider the yields of all the Ne isotopes: $^{20}$Ne, $^{21}$Ne, and $^{22}$Ne. It is worth recalling that  $^{20}$Ne is the most abundant isotope, and it is not significantly modified by nucleosynthesis during the AGB phase, but two $\alpha$-captures may transform a $^{14}$N into a $^{22}$Ne that could be mixed to the surface in the third dredge-up episode, increasing the total Ne abundance. Interestingly, the models predict large Ne increases at very low metallicities. $^{22}$Ne is efficiently produced and brought to the surface in stars  with initial masses between $\sim$2 and 4  M$_\odot$, depending on the metallicity. Stars with higher masses are less efficient because $^{22}$Ne diminishes as a consequence of an $\alpha$ capture that destroys it. 
Our objects are delimited by models with $Z$ between 0.001 and 0.004. In this case we found that  models predict excess of Ne, unless our objects have had  initial masses lower than 1.5 or 2 M$_\odot$, which contradicts the results from the N/O vs. O/H diagram using the same set of models (see Fig.~\ref{N-O} in previous section). 
In \S~\ref{models} this subject is discussed more deeply.

\medskip

%12+log(Ne/H)=(-0.176$\pm$0.566) + (0.925$\pm$0.071)*(12+log(O/H))
%\medskip
%r(Spearman)=0.88, r(Pearson)=0.45

\begin{figure}
\includegraphics[width=\columnwidth]{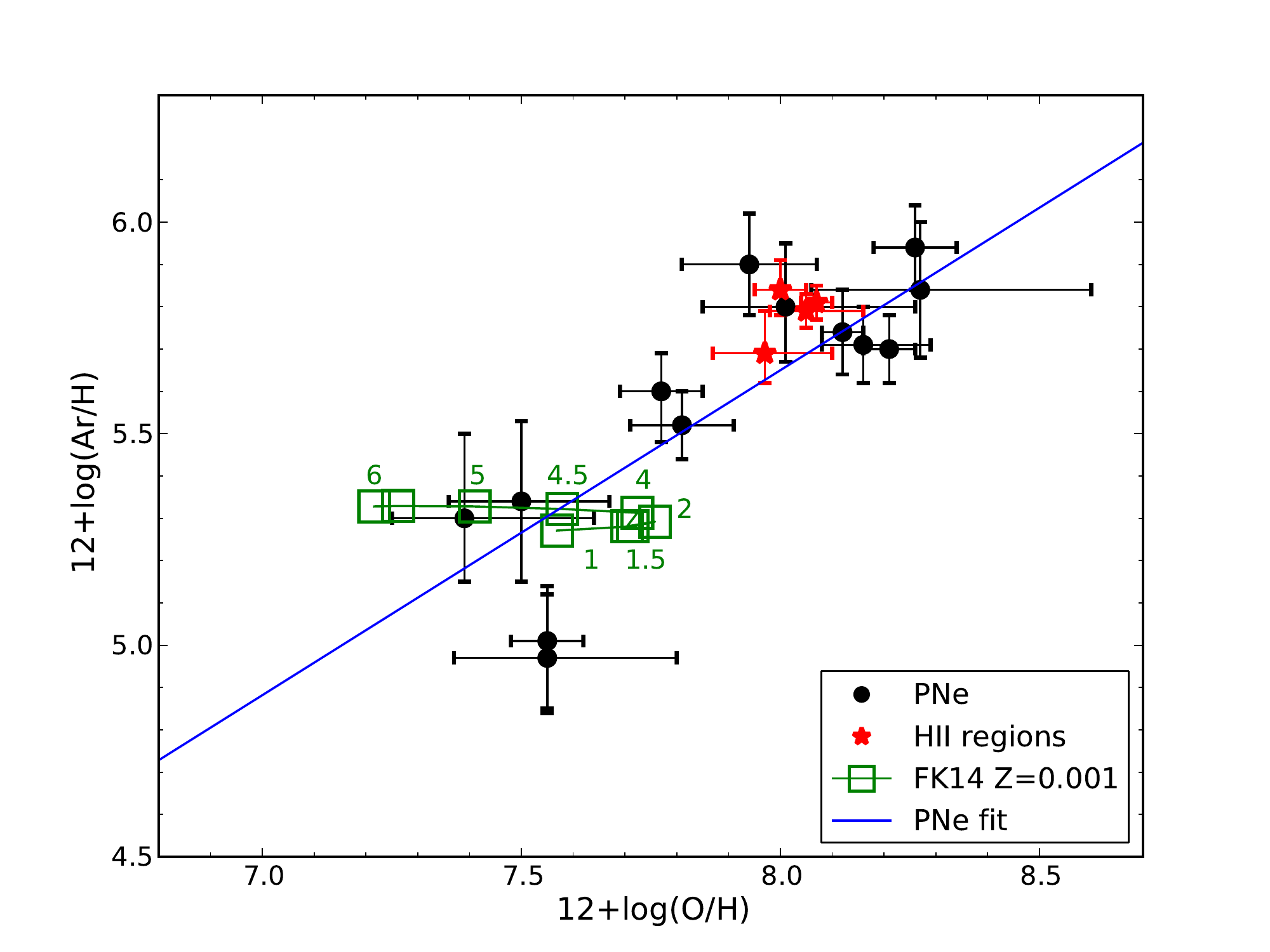}
\caption{Behaviour of 12+log (Ar/H) vs. 12+log (O/H). In red the {\hii} regions. The linear fits, including only PNe, corresponds to 12+log (Ar/H)=($-$0.494$\pm$0.512) + (0.768$\pm$0.064)$\times$(12+log (O/H)), with r= 0.79. The stellar evolution models for $Z$=0.001 by \citet[][F14]{fishlocketal14} are included. }
\label{Ar-O}
\end{figure}

The behaviour of Ar/H vs. O/H is presented  Fig~\ref{Ar-O}. A correlation is also found for these elemental abundances. Although it is not as tight as in the case of Ne/H vs. O/H, it shows that O, Ne, and Ar are elements that evolve closely in lock-steps. Data for {\hii} regions have been included in this diagram (red dots). In this case the slope of the correlation (computed including PNe only) is 0.768$\pm$0.064, significantly far from 1, which could indicate that O abundances are too large at the high-metallicity end, as would occur if O is enriched by stellar nucleosynthesis. 

The predictions of evolution models by \citet{fishlocketal14} at $Z$=0.001 are included in the graph. These models show clearly that Ar is not modified by stellar nucleosynthesis, independently of the initial stellar mass. In this case, the 12+log (Ar/H) abundance has an almost constant value around 5.3.

%12+log(Ar/H)=(-0.494$\pm$0.512) + (0.768$\pm$0.064) $\times$(12+log(O/H))
%r(Spearman)=0.79, r(Pearson)=0.40
\medskip
\begin{figure}
\includegraphics[width=\columnwidth]{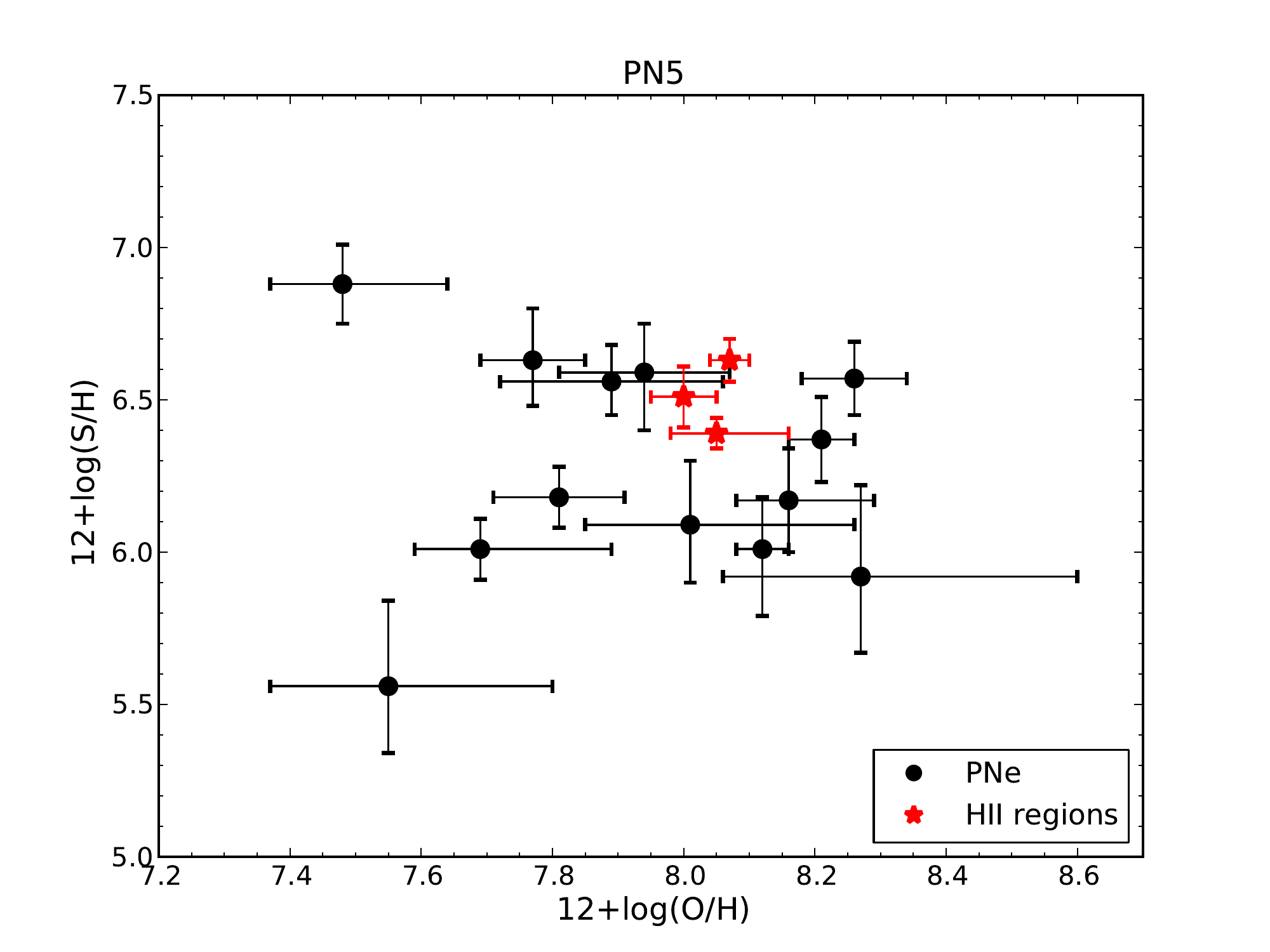}
\caption{Behaviour of 12+log (S/H) vs. 12+log (O/H), for PNe in NGC\,6822. In red the {\hii} regions. No correlation exists. The {\hii} regions present larger S abundances than many PNe.}
\label{S-O}
\end{figure}

S/H vs. O/H abundances are presented in Fig.~\ref{S-O}. In this graph no correlation of S with O is found, although it should be expected, because S is also an $\alpha$ element produced in similar  processes to O, Ne, and Ar. Probably this has to do with the difficulty of deriving the S total abundance for which no appropriate ICF exists \citep{henryetal04}. In this graph we see that S/H values in PNe spread in a wide range,  and it is also seen that {\hii} regions show a S/H ratio similar to many PNe, but an important fraction of PNe present lower S/H than the values in {\hii} regions. Therefore the phenomenon known as ``sulphur anomaly''  (consisting  in PNe showing sustantially lower S abundance than {\hii} regions at a given metallicity) is present in our PN sample. The reasons behind this phenomenon are still not clear, although as said above, it has been attributed mainly to deficient ICFs for calculating S abundance, particularly, to the uncertain contribution of S$^{3+}$ to the total S. In our case, the use of the new ICFs proposed by \citet{delgadoingladaetal14} has not improved the situation. On the other hand, some authors claim that the sulphur anomaly can be explained by the depletion of S into dust, especially in C-rich dust PNe \citep{garciahdezgorny14}. Computing abundances from S lines in IR spectra, where $S^{3+}$ lines can be measured, would shed some light on this problem.

\setcounter{table}{6}
\begin{table*}
\begin{tiny}
\begin{center}
\caption{Total abundances$^{\rm a}$ and L(H$\beta$) for PNe from VLT and CFH data and for H\,V and H\,X$^{\rm b}$.}
\label{total_ab}
\begin{tabular}{lcccccccc}
\noalign{\smallskip} \noalign{\smallskip} \noalign{\hrule} \noalign{\smallskip}
            Elem/object &        PN\,4          &          PN\,5                        &       PN\,6             &        PN\,7          &        PN\,10                 &        PN\,12           &       PN\,14     &      PN\,16     \\
\noalign{\smallskip} \noalign{\hrule} \noalign{\smallskip}
                 He  &10.92$\pm$0.04            & 11.03$\pm$0.10        & 10.99$\pm$0.04  & 11.09$\pm$0.08        & 11.16$\pm$0.03                & 11.01$\pm$0.05  & 11.02$\pm$0.04 &11.13$\pm$0.04        \\
                  N  & 7.43$\pm$0.09            & 8.21$^{+0.17}_{-0.12}$        & 7.55$\pm$0.18   & 6.89$^{+0.18}_{-0.16}$& 6.37$^{+0.13}_{-0.15}$& 7.50$\pm$0.17 & 8.39$\pm$0.12 & 8.13$\pm$0.12   \\
                  O  & 7.81$\pm$0.10            & 7.94$\pm$0.13         & 8.16$^{+0.13}_{-0.08}$& 8.27$^{+0.33}_{-0.21}$& 7.55$\pm$0.07   & 8.26$\pm$0.08 & 8.12$\pm$0.04& 8.21$\pm$0.05    \\
                Ne  & 7.00$\pm$0.08             &  \nodata                              & 7.65:                   & \nodata                       & 6.75$^{+0.20}_{-0.11}$& 7.42$\pm$0.08   & \nodata& 7.47$\pm$0.07                \\
                  S  & 6.18$\pm$0.10            & 6.59$^{+0.16}_{-0.19}$        & 6.17$\pm$0.17   & 5.92$^{+0.30}_{-0.25}$& \nodata       & 6.57$\pm$0.12 & 6.01$^{+0.17}_{-0.22}$& 6.37$\pm$0.14       \\
                Ar  & 5.52$\pm$0.08             &  5.90$\pm$0.12                                & 5.71$\pm$0.09   & 5.84$\pm$0.16& 5.01$^{+0.11}_{-0.16}$& 5.94$\pm$0.10  & 5.74$\pm$0.10 & 5.70$\pm$0.08           \\
\noalign{\smallskip} \noalign{\smallskip} \noalign{\hrule} \noalign{\smallskip} 
L(H$\beta$)/L$_\odot$ & 33.92  &  3.27 & 20.16 & 7.32 & 26.95 & 3.18 & 15.98 & 5.73 \\
\hline
\noalign{\smallskip} \noalign{\hrule} \noalign{\smallskip}
                 Elem/object &      PN\,18                      &     PN\,19                         &         PN\,21                &        {\hii}\,15 &         PN\,17 &     PN\,20   &  H\,V  &  H\,X  \\
\noalign{\smallskip} \noalign{\hrule} \noalign{\smallskip}
                 He  &11.09$\pm$0.09            &11.14$^{+0.14}_{-0.21}$        &11.11$^{+0.05}_{-0.09}$&10.95$\pm$0.03         & 11.12$^{+0.09}_{-0.11}$ & 10.96$^{+0.06}_{-0.07}$ &  10.94$\pm$0.02  & 10.95$\pm$0.03 \\
                  N  & 8.04$^{+0.14}_{-0.12}$   & 6.50$^{+0.33}_{-0.11}$        & 7.26$^{+0.17}_{-0.13}$& 6.54$\pm$0.04   & 6.62$^{+0.15}_{-0.14}$ & 6.33$^{+0.21}_{-0.33}$ & 6.72$\pm$0.05  & 6.63$\pm$0.06 \\
                  O  & 7.48$^{+0.16}_{-0.11}$   & 7.50$^{+0.17}_{-0.14}$        & 7.77$\pm$0.08   & 8.05$^{+0.11}_{-0.07}$        & 7.89$\pm$0.16 & 8.01$\pm$0.03 &  8.07$\pm$0.03 &  8.00$\pm$0.05 \\
                Ne  & 6.88$^{+0.19}_{-0.16}$    & \nodata                               & 7.00$^{+0.21}_{-0.16}$& 7.18$\pm$0.04 & 7.07$^{+0.24}_{-0.21}$ & 7.42$^{+0.11}_{-0.12}$ & 7.31$\pm$0.04 & 7.21$\pm$0.07\\
                  S  & 6.88$\pm$0.13                    & 6.10:                         & 6.63$^{+0.17}_{-0.15}$& 6.39$\pm$0.05           & 6.56$^{+0.12}_{-0.11}$ & 6.04$^{+0.17}_{-0.16}$ &  6.63$\pm$0.07 &  6.51$\pm$0.10 \\
                 Cl  &  \nodata                                 &  \nodata                                 &  \nodata                       &  4.50$\pm$0.09               & \nodata         & \nodata &  4.86$^{+0.10}_{-0.08}$  & 4.60$\pm$0.14 \\
                Ar  & \nodata                           & 5.34$\pm$0.19                 & 5.60$^{+0.09}_{-0.12}$ & 5.79$\pm$0.04          & \nodata & 5.80$\pm$0.06  & 5.81$\pm$0.03 & 5.84$\pm$0.06 \\
\noalign{\smallskip} \noalign{\smallskip} \noalign{\hrule} \noalign{\smallskip} 
L(H$\beta$)/L$_\odot$ & 0.57 &  1.23 & 0.76 & 308.67 & \nodata &  \nodata & \nodata & \nodata \\
\noalign{\smallskip} \noalign{\hrule} \noalign{\smallskip}
\end{tabular}
\end{center}
\begin{description}
\item[$^{\rm a}$] Abundances in 12+log (X/H).
\item[$^{\rm b}$] Colons indicate errors larger than a factor of two.
\end{description}
\end{tiny}
\end{table*}

\setcounter{table}{7}
\begin{table*}
\begin{tiny}
\begin{center}
\caption{Total abundances$^{\rm a}$ in NGC\,6822 from GTC data$^{\rm b}$.}
\label{total_ab_2}
\begin{tabular}{lcccccccc}
\noalign{\smallskip} \noalign{\smallskip} \noalign{\hrule} \noalign{\smallskip}
            Elem/object &        PN\,2          &          PN\,8                        &       PN\,11                    &        PN\,13                 &        PN\,23                  &        PN\,24         &      H\,II & H\,III\\
\noalign{\smallskip} \noalign{\hrule} \noalign{\smallskip}
                 He  &11.27$\pm$0.12            &11.01$\pm$0.05                 & \nodata                                 &11.13$\pm$0.06         &11.01$^{+0.09}_{-0.13}$        & $<$11.30         & $>$11.02 & \nodata \\
                  N  & 6.58$^{+0.21}_{-0.17}$& 7.28$\pm$0.32            & 8.27$^{+0.27}_{-0.14}$  & 6.39$^{+0.16}_{-0.23}$&  $>$7.38& \nodata             & 6.30: &         6.38$\pm$0.08 \\
                  O  & 7.55$^{+0.25}_{-0.18}$& 7.39$^{+0.25}_{-0.14}$   & 7.74:                           & 7.69$^{+0.20}_{-0.10}$        &  $>$8.23         & $>$ 7.83              & $>$7.63 & 7.97$^{+0.13}_{-0.10}$      \\
                Ne  & 6.76:                             & $<$6.42                               & \nodata                                 & \nodata                       & \nodata                         & \nodata               & \nodata & 6.88$^{+0.15}_{-0.12}$      \\
                  S  & 5.56$^{+0.28}_{-0.22}$ & 5.99:                           & \nodata                                 & 6.01$\pm$0.10 &  $>$6.49                      & $>$ 5.28                & \nodata &  \nodata    \\
                Ar  & 4.97$^{+0.17}_{-0.13}$    & 5.30$^{+0.20}_{-0.15}$        & 6.50:                           & $<$5.40                       &  $>$5.74                 &$>$ 5.51               & 5.48: &  5.69$^{+0.10}_{-0.07}$    \\
\noalign{\smallskip} \noalign{\smallskip} \noalign{\hrule} \noalign{\smallskip}
L(H$\beta$)/L$_\odot$ & 13.95 &  21.79 & 0.22 & 56.85 & 3.27 &  7.07 &\nodata \\
\noalign{\smallskip} \noalign{\hrule} \noalign{\smallskip}
\end{tabular}
\end{center}
\begin{description}
\item[$^{\rm a}$] Abundances in 12+log (X/H).
\item[$^{\rm b}$] Colons indicate errors larger than a factor of two.
\end{description}
\end{tiny}
\end{table*}

\subsection{Testing other stellar evolution models
\label{models}}

As said above, the comparison of the observed N/O abundances with predictions of stellar evolution models by \citet{karakas10} and \citet{fishlocketal14} indicates that the initial masses of the observed PNe were lower than 4 M$_\odot$, contradicting the predictions of the same models for Ne/H vs. O/H (Fig.~\ref{Ne-O}),  from which the initial masses should have been lower than 2 M$_\odot$. Therefore we have decided to analyse the behaviour of models more deeply.

\citet{karakaslattanzio03} have compared the Ne yield of their models with abundances of PNe of the Milky Way ($Z$ = 0.02) and the Large Magellanic Cloud  ($Z$ = 0.008) to find that models agree with the observations except for the case of models with  initial mass of 3 M$_\odot$ and $Z$=0.008, which  seem to predict too much Ne. 

From the analysis of a sample of Galactic disk and bulge double chemistry (DC), O-rich dust (OC), and C-rich dust (CC) PNe, \citet{garciahdezgorny14} found that the average Ne/O at $Z = 0.02$ for their DC and OC PNe was slightly larger than those predicted by the models by \citet{karakas10} without a partial mixed zone (PMZ); therefore, they conclude that including and varying the size of a PMZ can solve the observed behaviour. However, that is exactly the opposite of what we found in our sample. For our objects (which have $Z~<$ 0.004), we found that models with initial mass higher than 2 M$_\odot$ show too much Ne, in comparison with observations. We consider that it is possible that such models are producing an excess of $^{22}$Ne or that the efficiency of the third dredge-up should be diminished in those stars. 

Very recently, \citet{venturaetal15} and \citet{dellaglietal15} have used a new generation of AGB stellar models that include dust formation in the stellar winds to constrain CNO abundances of PNe in the Magellanic Clouds, which have a metallicity slightly higher than but comparable to that of NGC\,6822. These models were developed using the ATON evolutionary code, which stands apart from others in that it uses the \citet{blocker95} mass loss prescription and the full spectrum of turbulence (FST) convective model \citep{canutomazzitelli91}. These assumptions concern the overshoot of the convective core during the core H-burning phase and lead to a less efficient dredge-up and to a lower threshold mass for the activation of the HBB. A detailed description of these models are given in \citet{venturaetal14} for $Z=0.004$, in \citet{venturaetal13} for $Z=0.001$ and M $>$ 3 M$_{\odot}$, and in \citet{venturaetal14b} for $Z=0.001$ and M $\leq$ 3 M$_{\odot}$. 

\begin{figure}
\includegraphics[width=\columnwidth]{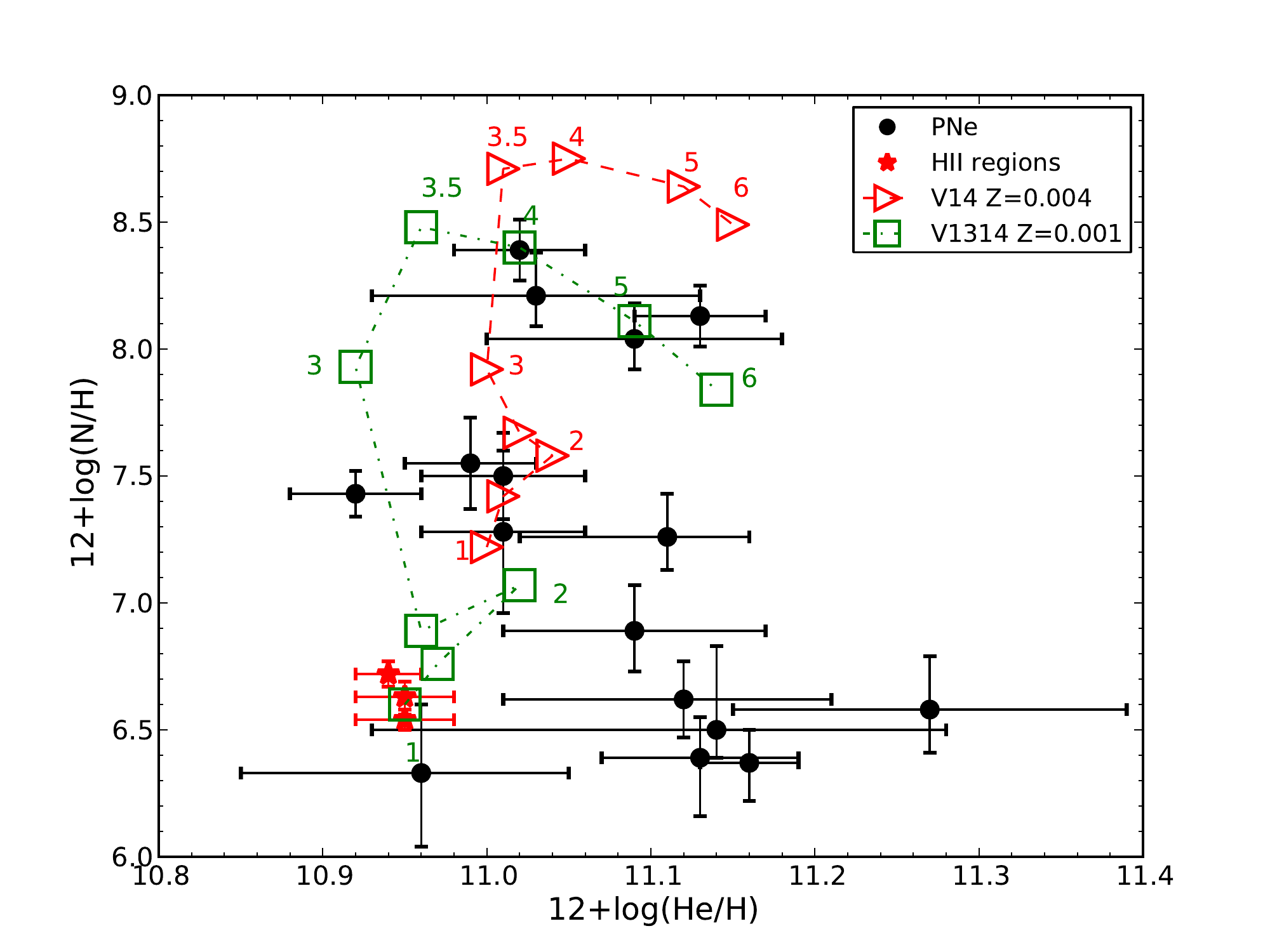}
\includegraphics[width=\columnwidth]{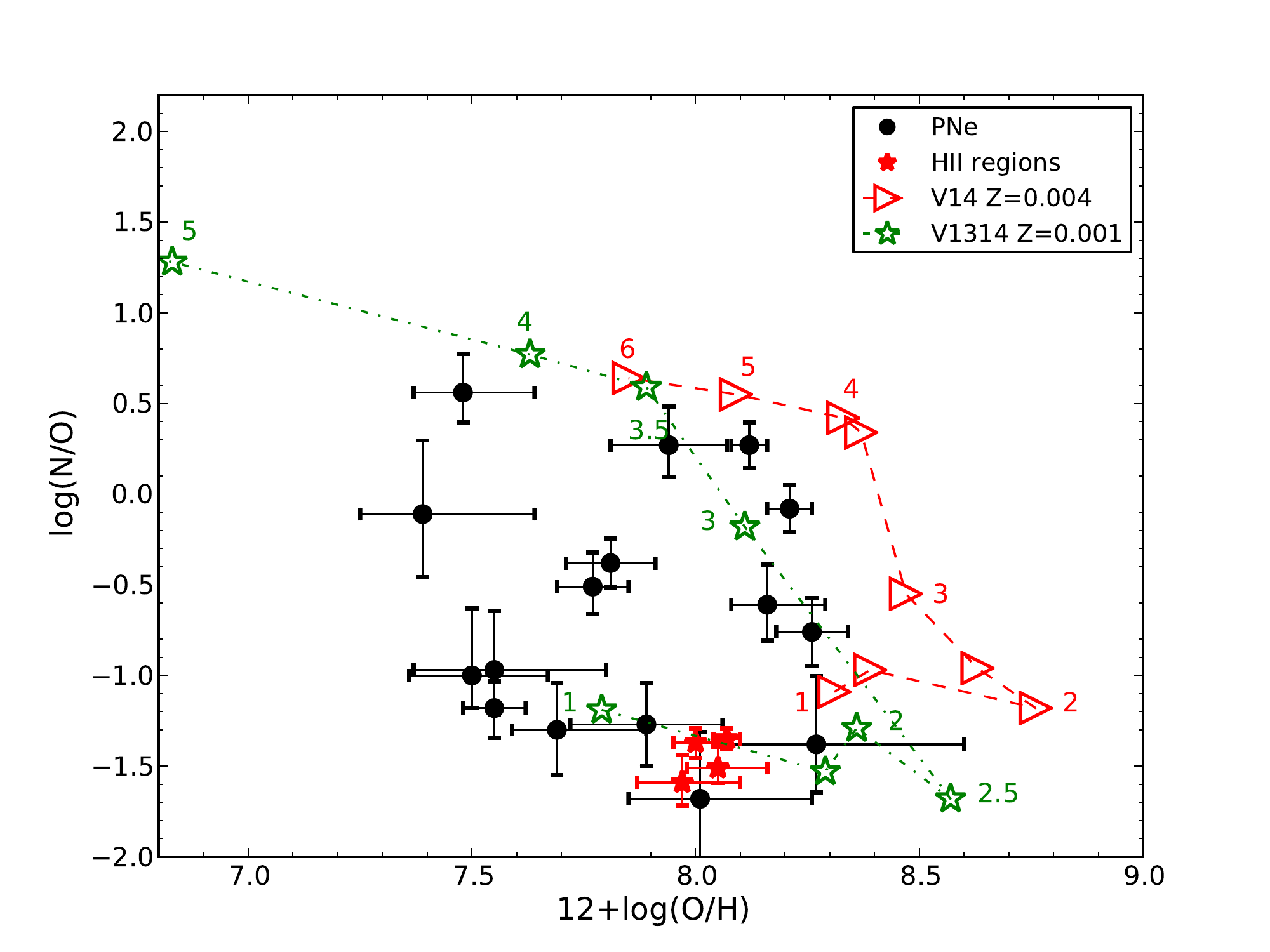}
\caption{ Same as Fig.~\ref{N-O} but showing the predictions of stellar evolution models for $Z=0.001$ (green squares) by \citet[][V1314]{venturaetal13, venturaetal14b} and $Z=0.004$ (red triangles) by \citet[][V14]{venturaetal14}. See text for discussion.  }
\label{N-Oalt}
\end{figure}
 
\begin{figure}
\includegraphics[width=\columnwidth]{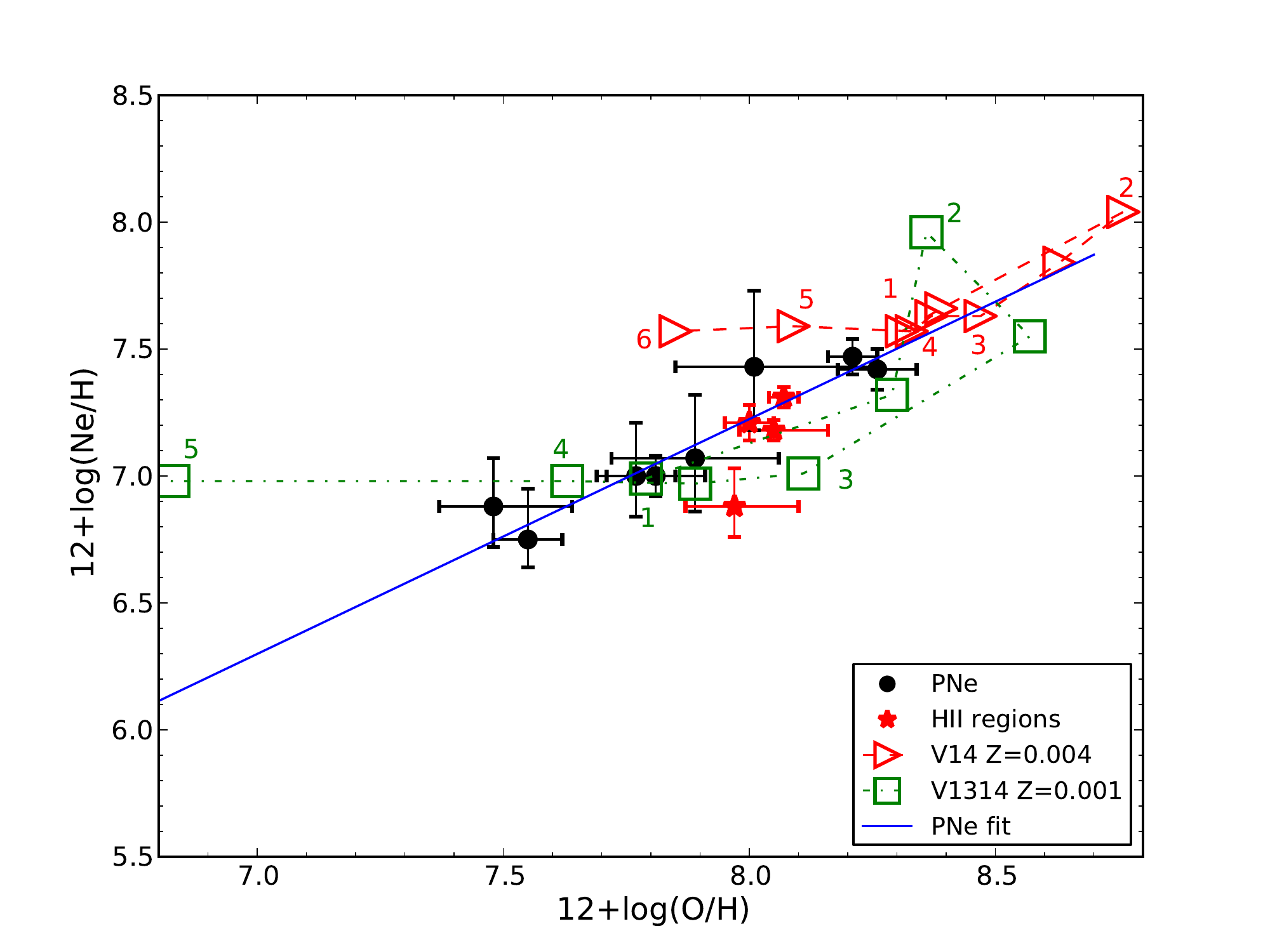}
\caption{Same as Fig.~\ref{Ne-O} but showing the predictions of stellar evolution models for $Z=0.001$ (green squares) by \citet[][V1314]{venturaetal13, venturaetal14b} and $Z=0.004$ (red triangles) by \citet[][V14]{venturaetal14}. See text for discussion.}
\label{Ne-Oalt}
\end{figure}

In Figs.~\ref{N-Oalt} and~\ref{Ne-Oalt} we explore the behaviour of 12+log (N/H) vs. 12+log (He/H) and log (N/O), and 12+log (Ne/H) vs. 12+log (O/H) with models by \citet{venturaetal13, venturaetal14b} for $Z$=0.001 and \citet{venturaetal14} for $Z$=0.004. It is found that they predict lower N/O ratios for intermediate-mass progenitors stars than the models by \citet{karakas10} and \citet{fishlocketal14}.  Additionally, the lowest N/O ratios are reproduced by stars with initial masses lower than 2.5  M$_\odot$. It is important to stress than in the $Z$=0.001 models, the HBB destroys O very efficiently in the more massive progenitors.  In our sample there are several objects at 12+log (O/H) $<$ 7.8 that are not reproduced by the set of models shown here. Probably these objects could be reproduced by models at $Z$=0.0001. In Fig.~\ref{N-Oalt} (above), it is observed that Ventura et al. models do not reproduce the large He/H abundance found in the PNe with 12+log(N/H) $<$ 7.0).

Similar to what it was found for N, in Fig.~\ref{Ne-Oalt} the observed Ne abundances are consistent with models of stellar progenitors with masses up to 4-5 M$_\odot$, which is in overall agreement with what we found in Fig.~\ref{N-Oalt}. This is because dredge-up events   in \citet{venturaetal15} models are less efficient than the Karakas ones. 

\begin{figure}
\includegraphics[width=\columnwidth]{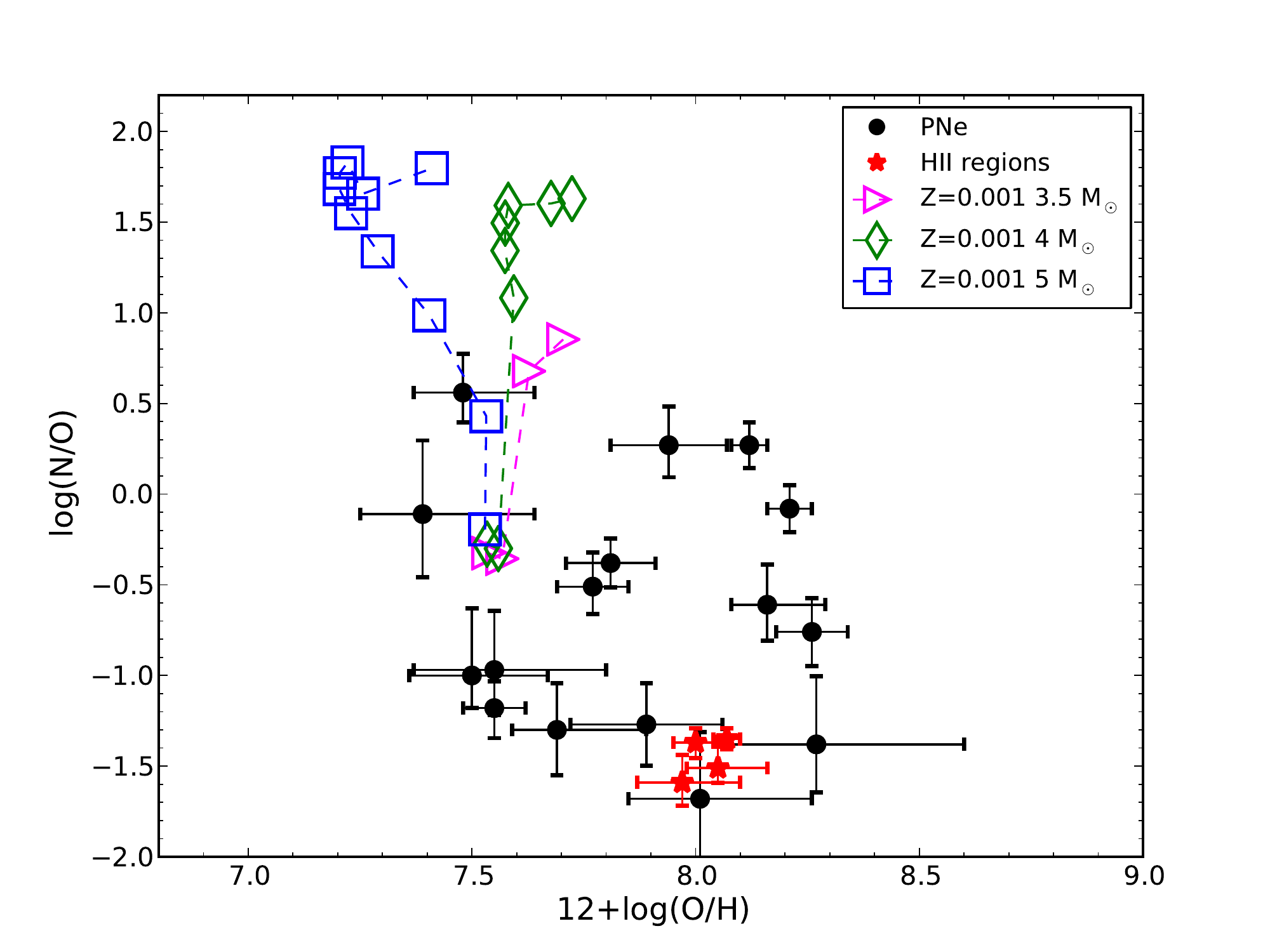}
\includegraphics[width=\columnwidth]{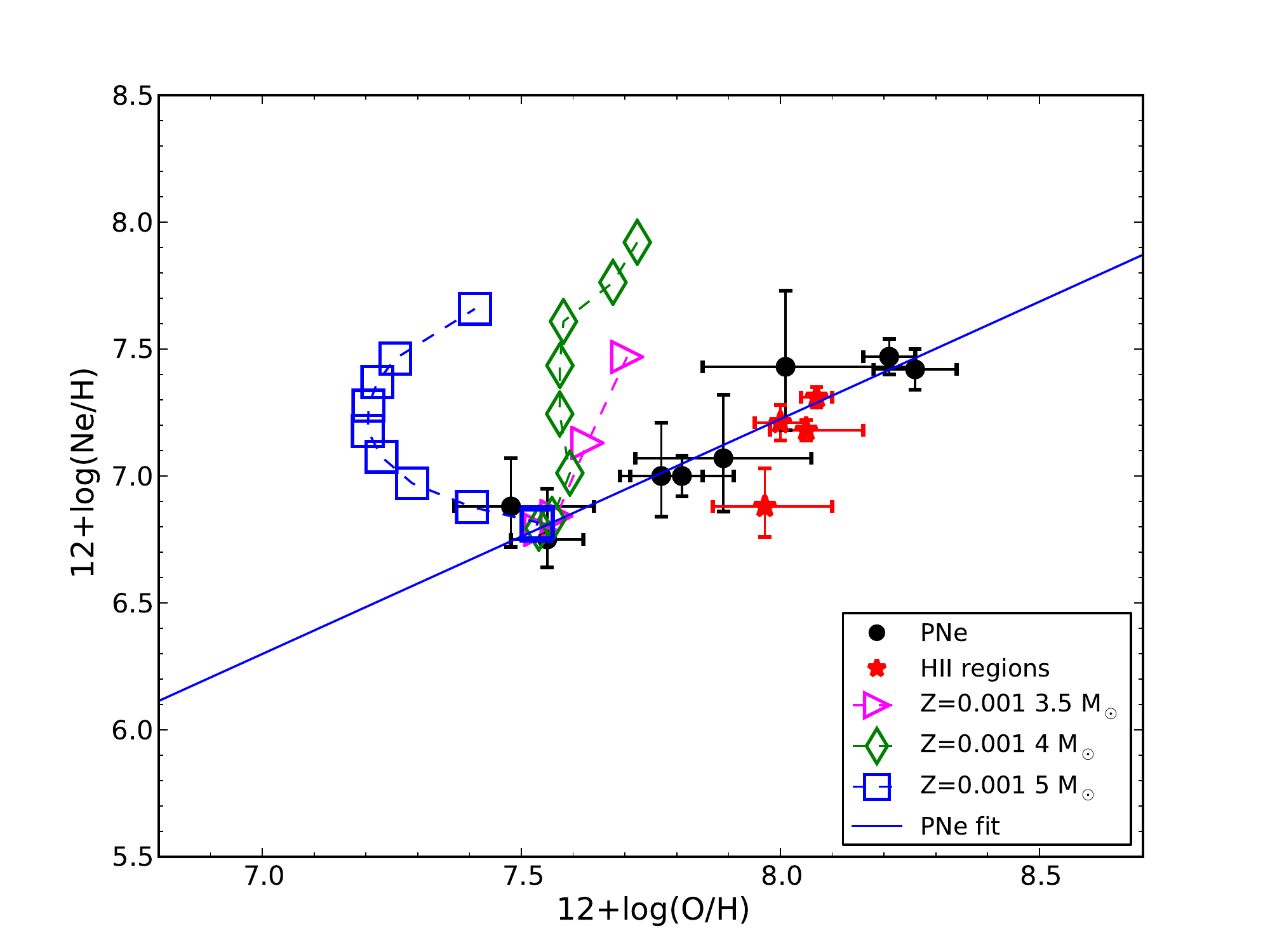}
\caption{Evolution of the surface abundances for models with $Z$=0.001 and initial masses of 3.5, 4, and 5 M$_{\odot}$ of \citet{fishlocketal14}. First point shows abundances after first TP and the last point represents the final surface abundances of the AGB model. The step between points is  $\sim$10 TP.}
\label{steps}
\end{figure}

On the other hand, we checked the evolution of the surface abundances obtained from the \citet{fishlocketal14} models during the different thermal pulses (TP) and dredge-up events. In Fig.~\ref{steps} we show the behaviour of the N/O and Ne/H surface abundances for models with initial masses of 3.5, 4, and 5 M$_{\odot}$ at different steps, from the first TP to the final  surface abundances in steps of 10 TP, and we found that if the models are stopped significantly before (the amount of mass loss changes significantly) they can also reproduce the observed behaviour in N/O.

A deeper discussion of this subject is beyond the scope of this paper, but it would be important to analyse the predictions of models in comparison with the chemical abundances  of different elements (C, N, O, Ne, Ar, etc.)  observed in PNe at different metallicities, looking for a good agreement between observations and theory.

\subsection{PNe populations}

From a sample of 11 objects \citet{hdezmartinezetal09} reported the presence of two PN populations, one older, of low metallicity with 12+log (O/H) values lower than 7.9, and one younger, of high metallicity, including PNe showing 12+log (O/H) abundances larger than 8.0 and similar to the {\hii} region abundances (the four {\hii} regions presented in Tables 7 and 8 show a very similar 12+log (O/H) value with an average of 8.02$\pm$0.05). In our extended PN sample, such a segregation does not seem so evident. However, there is only one PN with 12+log (O/H) in the range from 7.9 to  8.0, therefore our sample still supports the idea of having two PN populations on the basis of O/H abundances.

By analysing the behaviour of Ar abundances in Fig~\ref{Ar-O}, two PN populations can also be differentiated, one showing 12+log (Ar/H) lower than 5.7 and the other Ar/H similar to the values in {\hii} regions, which on average is 5.80$\pm$0.10.

As said in the Introduction, \citet{carigietal06} used \citet{wyder03} data and proposed a star formation history  representative of the whole galaxy. In their Fig. 4 presenting the star formation rate (SFR), a clear burst appears between 6 and 8 Gyr, followed by a fall in the SFR lasting up to about 10 Gyr, when a second star formation episode starts. These two episodes of star formation coincide with the ages assigned by \citet{hdezmartinezetal11} to the PN two populations, between 3 and 9 Gyr for 
the older one and   ages under 3 Gyr  for the younger one. These PN ages were obtained from \citet[][and references therein]{allenetal98}. Additionally, according to \citet{rguezglezetal15}, the presence of a dip in the planetary nebulae luminosity function (PNLF) in NGC\,6822 indicates that  two important episodes of star formation should have occurred in order to obtain such a dip in the PNLF. Therefore, our finding of two PN populations in NGC\,6822 and their ages corroborate the two star formation episodes found in the SFR by \citet{carigietal06}. 

Regarding the Ar abundances, as expected within uncertainties, no PN shows Ar/H larger than {\hii} regions. However this is not the case of O, for which a few PNe have O/H slightly larger than {\hii} regions (see Fig~\ref{Ar-O}). Similar to the slope of the Ar/H vs. O/H correlation, this could indicate O-enrichment in the PN shell due to nucleosynthesis in the stellar interior. Such an enrichment has been predicted by some stellar evolution models for low-metallicity stars \citep[see e.g.][]{marigo01} and from some nucleosynthesis models including extra mixing mechanism \citep{pignatarietal13}. 
\citet{penaetal07} find O-enrichment in the PNe of the low-metallicity irregular galaxy NGC\,3109. Such an O-enrichment has also been found in some PNe of the Milky Way showing C-rich dust \citep{delgadoingladaetal15}, and it is attributed to an efficient third dredge-up episode. In case of O-enrichment, O abundance does not trace the metallicities of objects and, therefore Ar or other elements, such as Cl, are better indicators of initial metallicities.

\section{Results and discussion
\label{sec:discuss}} 

From observations obtained with the GTC, VLT, and CFH telescopes, we have analysed the spectrophotometric data of 22 PNe in NGC\,6822. This represents 84\% of the total sample of 26 PN candidates known in this galaxy. We have confirmed the PN nature of 73\% of the sample. Physical conditions and chemical abundances (in particular helium, nitrogen, oxygen, argon, and sulphur) were  derived 
for 19 PNe. In addition, the chemical abundances of four {\hii} regions have been included for comparison: {\hii}\,15
and H\,III observed by us and H\,V and H\,X from data by \citet{peimbertetal05}. In the following we discuss the main results of our analysis.

A large number of Peimbert Type I PNe were found. \citep[Following][ Peimbert Type I PNe were selected as those having N/O abundance ratio larger than 0.5.]{hdezmartinezetal09} These are  PN\,5, PN\,8, PN\,11, PN\,14, PN\,16, and PN\,18, representing  over 31\%
of the analysed sample (very probably PN\,25 is also a Type I PN; but it is not included here since we cannot calculate its chemical abundances, and it is potentially affected by atmospheric diffraction effects, see Sect.~\ref{sec:obs}).  This large number resembles the number of Type I PNe found in the galactic bulge, M\,31 and the LMC  \citep{stasinskaetal98}. We recall that Type I PNe  are He- and N-rich objects and they originated in stars that are more massive than in other PN types. Therefore it seems that we are  detecting a large number of PNe produced by the more massive central stars, and the same is true for other galaxies. From the comparison of our N/O values and the predictions of stellar evolution models (Figs.~\ref{N-O} and~\ref{N-Oalt}), it is found that the initial masses of the central stars were lower than 4 M$_\odot$ and that Type I PNe had masses between 3 and 4 M$_\odot$.

\begin{figure}[!th]
\begin{center}
\includegraphics[width=\columnwidth]{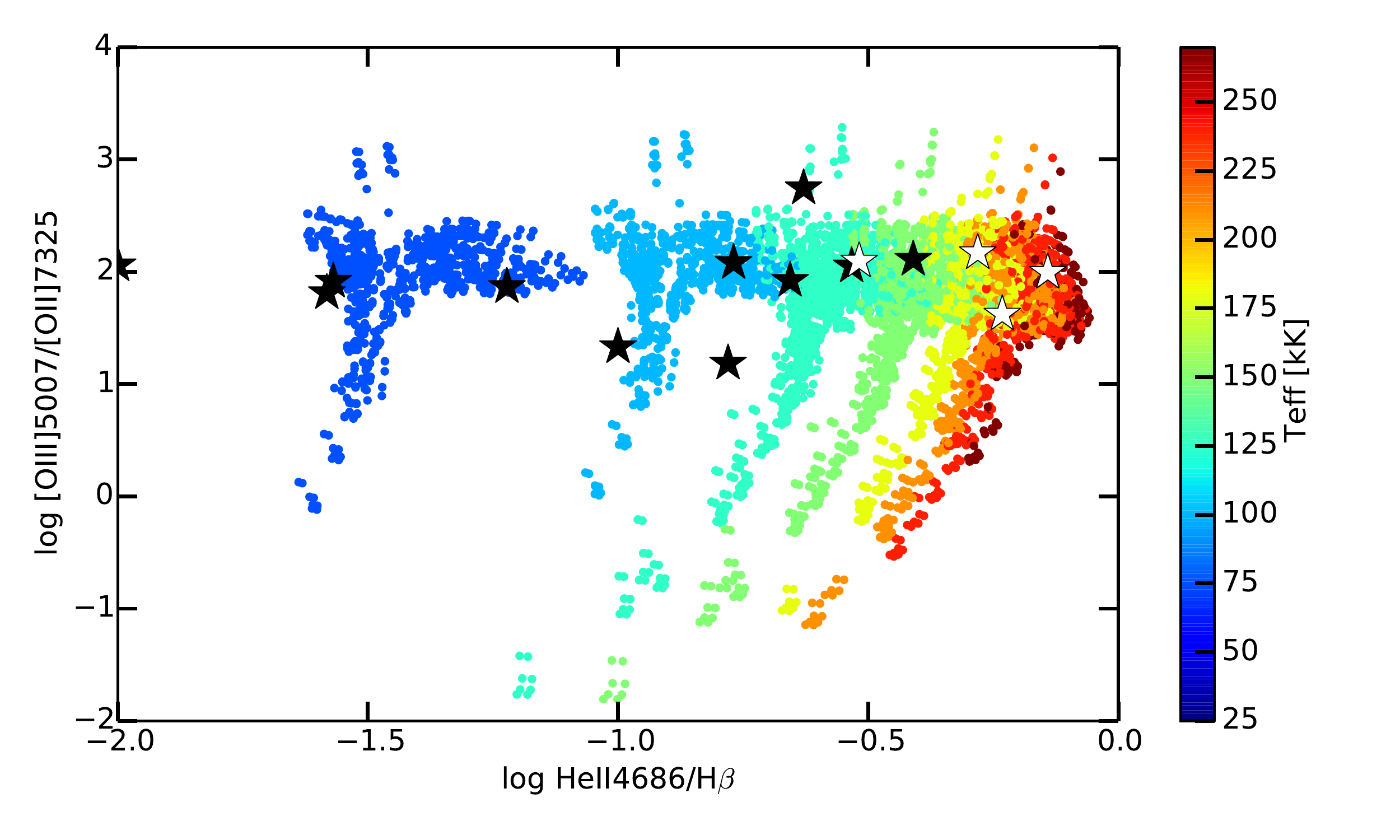}
\caption{$\lambda$5007/{\foii}$\lambda$7325 vs. {\heii} $\lambda$4686/H$\beta$ as obtained from the Mexican Million Models data base by \citet{morissetetal15}. Stellar temperatures can be obtained from this plot. Many of our PNe have stars with T higher than 10$^5$ K. The Type I PNe are indicated in white.\label{TeHeII}}
\end{center}
\end{figure}

Highly ionized PNe with a ionic fraction He$^{++}$/He $\geq$ 0.10 are PN\,4, PN\,5, PN\,8, PN\,12, PN\,14, PN\,16, PN\,23, PN\,24, PN\,13, PN\,18, and PN\,19, representing almost 60\% of the sample. 
This high percentage of He$^{++}$/He implies the presence of a very hot central star. Models from the ample grid computed by \citet{morissetetal15} (Mexican Million Models data base, 3MdB) have been used to analyse this. We chose the `Realistic' and matter-bounded  PNe models (PNE-2014, Delgado-Inglada et al. 2014) that use a black body (BB) for the spectral energy distribution of the ionizing star. From this grid we have constructed Fig.~\ref{TeHeII}, showing the behaviour of our objects in comparison with photoionization  models, in an {\foiii}$\lambda$5007/{\foii}$\lambda$7325 vs. {\heii} $\lambda$4686/H$\beta$ diagram. The effective temperature for the models is shown in the colour bar. Our sample of  PNe with He$^{++}$/He $\geq$ 0.10 corresponds to objects with effective temperatures over 100~000 K. It is important to notice that we have a limited number of objects with {\foii}$\lambda$7325, but this is not a problem because in this figure we can clearly see that the ratio {\heii} $\lambda$4684/H$_\beta$ is the most important parameter for determining the effective temperature,  which is almost independent of the {\foiii}$\lambda$5007/{\foii}$\lambda$7325 ratio. The large number of highly ionized PNe  in NGC\,6822 seems unusual when compared to the PNe in our galaxy, but it is similar to what is found in the LMC \citep{stasinskaetal98}. Type I PNe have been marked in green in this figure. It is evident that these PNe present the highest {\heii} $\lambda$4686/H$\beta$ line intensity ratios and the highest stellar temperatures, which are in the range from 140 to 230 kK. If we compare such temperatures with the evolutionary tracks presented by \citet{vassiliadiswood94} for the SMC metallicity (their Fig. 5), we conclude that the three hottest stars had initial masses higher than 2.5 M$_\odot$ (and would have present core masses higher than 0.69 M$_\odot$). This agrees with the initial masses derived for Type Is from the N/O ratio (Figs.~\ref{N-O} and~\ref{N-Oalt}).

Therefore our PN sample seems to have some biases:  We find too many Type I PNe and very hot central stars (as compared to the galactic values).  Our PN candidates  were selected from a sample of {\foiii}$\lambda$5007- and H$\alpha$- emitting objects \citep{hdezmartinezpena09}. Actually to be certain they are PN and to eliminate compact {\hii} regions, \citet{ciardulloetal02} suggested selecting PN candidates because they have  {\foiii} $\lambda$5007/H$\alpha$  ratios  over 1.6. In the sample by \citet{hdezmartinezpena09}, there are a few candidates with {\foiii} $\lambda$5007/H$\alpha$ lower than 1.6, since other criteria were used, such as the non-detection of the central star. Either way, possible  low-ionization PNe with faint {\foiii} $\lambda$5007 and therefore low stellar effective temperature (T$_*$ lower than 40 kK) were ignored.

\begin{figure*}[!th]
\begin{center}
\includegraphics[width=\textwidth]{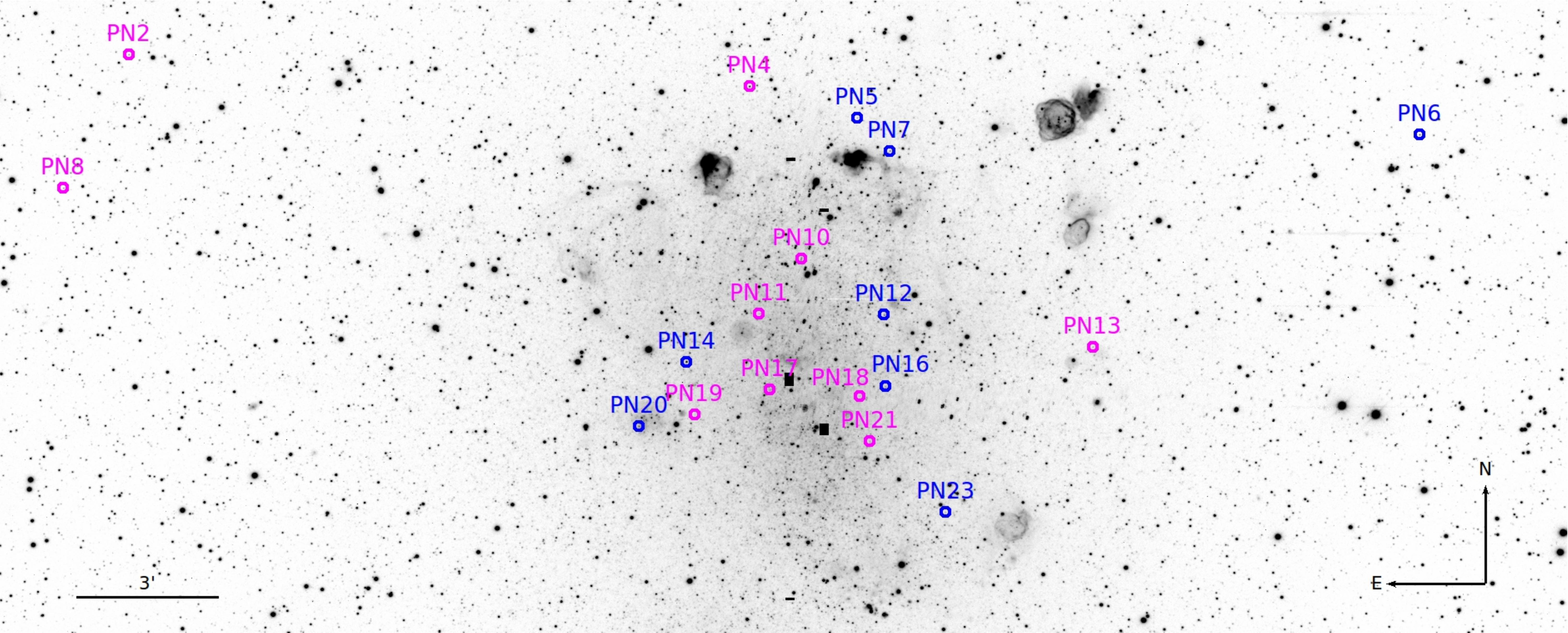}
\caption{Distribution of PNe in NGC 6822 in an H$\alpha$ image. In pink the objects with low metallicity (12+log (O/H) $<$ 8 and 12+log (Ar/H) $<$ 5.7); in blue, the high-metallicity ones (12+log (O/H) $>$ 8 and 12+log (Ar/H) $>$ 5.7.  PN\,6, with 12+log (Ar/H) of 5.71 is marginally a young object). \label{Ha-PN-ages}}
\end{center}
\end{figure*}

\begin{figure}[!th]
  \begin{center}
\includegraphics[width=\columnwidth]{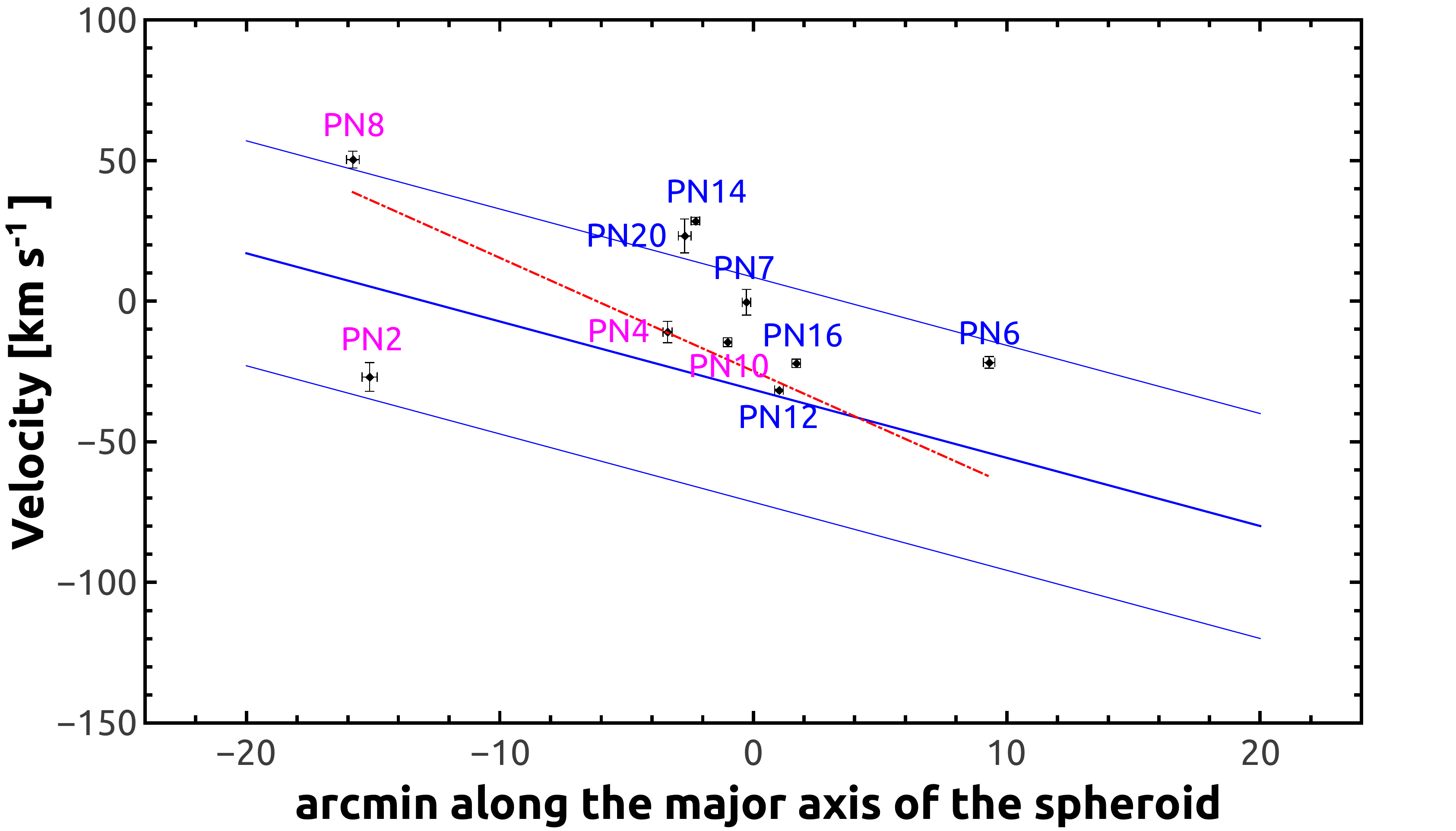}
  \caption{Position-velocity of PNe in NGC\,6822 with abundances determined, adapted from \citet{floresduranetal14}. The position is projected on the long axis of the stellar spheroid. In pink the old objects (12+log (Ar/H) $\leq$ 5.7); in blue the young objetcs.  \label{pos_vel}}
\end{center}
\end{figure}

As already reported for the 11 PNe studied by \citet{hdezmartinezetal09}, here we again find an old population of PNe with 12+log (O/H) $<$ 8 and 12+log (Ar/H) $<$  5.7 (see Figs.~\ref{L-O} and ~\ref{Ar-O}). These old objects are PN\,2, PN\,4, PN\,8, PN\,10, PN\,11, PN\,13, PN\,18, PN\,19, PN\,21, and PN\,17,  representing half of the total sample. Seven PNe show 12+log (O/H) greater than 8.0 and 12+log (Ar/H) over 5.7, and these  young PNe  are PN\,5, PN\,6, PN\,7, PN\,12, PN\,14, PN\,16, PN\,23, and  PN\,20. (It is worth noticing that PN\,6 is marginally young as its 12+ log (Ar/H) is 5.71.) In Fig.~\ref{Ha-PN-ages} we show the spatial distribution of the two populations in the galaxy. It is observed than young PNe tend to lie in the central zone well inside the optical bar, while the old objects are more widely distributed. The same is true if we plot the PNe velocities (relative to the system) against the distance along the long axis of the stellar spheroid, presented in Fig.~\ref{pos_vel} \citep[adapted from][]{floresduranetal14}; although no preference in velocity is found, the old population is distributed at both sides of the centre, while the young PNe are nearer the central region in the optical bar. We notice that PN\,6 could be marginally an old object.

\medskip
\section{Summary
\label{sec:summary}} 

The PN sample analysed here (representing  $\sim$84\% of the total sample detected in this galaxy) is biased towards  nebulae with very hot central stars. Such a bias probably occurs also in the PN samples of many other external galaxies due to the way PN candidates are selected. 

From comparison with stellar evolution models by \citet{karakas10} and \citet{fishlocketal14} of the N/O abundance ratio, our PNe should have had initial masses lower than 4 M$_\odot$, although if the comparison is made with Ne/H vs. O/H abundances, the initial masses  should have been lower than 2 M$_\odot$. It appears that the models of stars of  2$-$3 M$_\odot$ by \citet{karakas10} and \citet{fishlocketal14} are producing too much $^{22}$Ne in the stellar surface at the end of the AGB. On the other hand, the comparison with ATON models by \citet{venturaetal13, venturaetal14, venturaetal14b} -- which differ from the previous models in the treatment of convection and on the assumptions concerning the overshoot of the convective core during the core H-burning phase -- leads to reasonable agreement between the observed and predicted ratios of N/O and Ne/H if more massive stars are about 4 M$_{\odot}$. So far, none of the models reproduce the large He abundances found in many  PNe of NGC\,6822.

The Type I PNe were produced by stars of any metallicity (not necessarily the richer ones), and their initial masses were between 3 and 4 M$_\odot$. These objects show the highest effective temperatures.

The PNe in NGC\,6822 span in metallicity from very poor objects (12+log (Ar/H) $\leq$ 5.7) to nebulae showing the same metallicity as the {\hii} regions (12+log (Ar/H) $\sim$ 5.8). The poorer objects are more widely spread in the galaxy, while the young ones lie in or very near  the optical bar.

\begin{acknowledgements}

This work is based on observations with the Gran Telescopio Canarias (GTC), installed in the Spanish Observatorio del Roque de los Muchachos of the Instituto de Astrofísica de Canarias on the island of La Palma. This work has been funded by the Spanish Ministry of Economy and Competitiveness (MINECO) under the grant AYA2011-22614 and also received partial support from the DGAPA-UNAM, Mexico under grant PAPIIT IN109614. We thank the referee, G. Stasi\'nska, for her valuable comments that helped to improve the quality of the paper. JGR acknowledges support from the Severo Ochoa excellence programme (SEV-2011-0187) postdoctoral fellowship. We are very grateful to Dr. P. Ventura for providing the tables with the surface abundances of his AGB models, and to Dr. C. Morisset for his invaluable help and comments related to the 3Mdb. We also are grateful to A. Karakas from providing the detailed numbers of the models by \citet{fishlocketal14}. We acknowledge fruitful discussions with D.~A. Garc\'{\i}a-Hern\'andez, A. Karakas, and M. Lugaro.
\end{acknowledgements}

%-------------------------Bibliography--------------------------------------------

%-------------------------Online material-----------------------------------------

\Online

\end{document}